\documentclass[aps,pra,twocolumn,nofootinbib,floatfix,superscriptaddress]{revtex4-2}

\usepackage{amsmath,mleftright,mathtools}
\usepackage{bm}
\usepackage{stix,microtype}
\usepackage{graphicx}
\usepackage{xspace}
\usepackage{units}
\usepackage[colorlinks,linkcolor=blue,citecolor=blue,urlcolor=blue]{hyperref}
\usepackage[dvipsnames]{xcolor}
\usepackage{cleveref}
\usepackage{array}   
\newcolumntype{L}{>{$}c<{$}} 
\newcolumntype{C}{>{$}c<{$}} 
\usepackage{dcolumn}

\newcommand{\Hh}{\hat{H}}
\newcommand{\bos}{\text{bos}}
\newcommand{\el}{\text{el}}
\newcommand{\elbos}{\text{el-bos}}

\newcommand{\bt}{\bar{t}}

\newcommand{\hphi}{\hat{\phi}}

\newcommand{\hd}{\hat{d}}

\graphicspath{{./Figs/}}
\mathchardef\mhyphen="2D \newcommand{\nn}{\nonumber}

\newcommand{\be}{\begin{equation}}
\newcommand{\ee}{\end{equation}}

\newcommand{\mv}{\mathbfit{v}}
\newcommand{\mw}{\mathbfit{w}}
\newcommand{\mh}{\mathbfit{h}}

\newcommand{\mO}{\bm{\Omega}}
\newcommand{\mD}{\mathbfit{D}}

\newcommand{\mPsi}{\bm{\Psi}}

\newcommand{\mSgm}{\mathbfit{\Sigma}}

\newcommand{\mPi}{\bm{\Pi}}

\newcommand{\cG}{\mathcal{G}}

\newcommand{\mcG}{\bm{\mathcal{G}}}
\newcommand{\mI}{\mathbfit{I}}

\newcommand{\mP}{\mathbfit{P}}
\newcommand{\mrho}{\bm{\rho}}
\newcommand{\malpha}{\bm{\alpha}}
\newcommand{\bgg}{\bm{\gamma}}
\newcommand{\mchi}{\bm{\mathit{\chi}}}

\def\bcallC{\mbox{\boldmath $\mathcal{C}$}}
\def\bgm{\mbox{\boldmath $\mu$}}

\def\fbgm{\mbox{\scalebox{.7}{$\scriptscriptstyle \bgm$}}}

\def\bga{\mbox{\boldmath $\alpha$}}
\def\bgd{\mbox{\boldmath $\delta$}}
\def\bgg{\mbox{\boldmath $\gamma$}}
\def\bcallg{\mbox{\boldmath $g$}}
\def\bgx{\mbox{\boldmath $\chi$}}

\keywords{Nonequilibrium Green's function theory, generalized Kadanoff-Baym Ansatz, excited states}
\begin{document}
\title{Time-linear scaling NEGF methods for real-time simulations of 
interacting electrons and bosons. I. Formalism}
\author{Y. Pavlyukh}
\affiliation{Dipartimento di Fisica, Universit{\`a} di Roma Tor Vergata, Via della Ricerca Scientifica 1,
00133 Rome, Italy}
\author{E. Perfetto}
\affiliation{Dipartimento di Fisica, Universit{\`a} di Roma Tor Vergata, Via della Ricerca Scientifica 1,
00133 Rome, Italy}
\affiliation{INFN, Sezione di Roma Tor Vergata, Via della Ricerca Scientifica 1, 00133 Rome, Italy}
\author{Daniel Karlsson}
\affiliation{Department of Physics, Nanoscience Center P.O.Box 35
FI-40014 University of Jyv\"{a}skyl\"{a}, Finland}
\author{Robert van Leeuwen}
\affiliation{Department of Physics, Nanoscience Center P.O.Box 35
FI-40014 University of Jyv\"{a}skyl\"{a}, Finland}
\author{G. Stefanucci}
\affiliation{Dipartimento di Fisica, Universit{\`a} di Roma Tor Vergata, Via della Ricerca Scientifica 1,
00133 Rome, Italy}
\affiliation{INFN, Sezione di Roma Tor Vergata, Via della Ricerca Scientifica 1, 00133 Rome, Italy}
\date{\today}
\begin{abstract}  
Simulations of interacting electrons and bosons out of equilibrium, starting from first
principles and aiming at realistic multiscale scenarios, is a grand theoretical challenge.
Here, using the formalism of nonequilibrium Green's functions
and relying in a crucial way on the recently discovered time-linear formulation of
the Kadanoff-Baym equations, we present a versatile toolbox for the simulation of 
correlated electron-boson dynamics. A large class of methods 
are available, from the Ehrenfest to the dressed $GD$  for the 
treatment of electron-boson interactions in combination with 
perturbative, i.e., Hartree-Fock and second-Born, or 
nonperturbative, i.e., $GW$ and $T$-matrices either without or with 
exchange effects, for the treatment of the Coulomb interaction.  
In all cases the numerical scaling is linear in 
time and the equations of motion satisfy all fundamental conservation 
laws.

\end{abstract}
\maketitle
\section{Introduction}                            
The electron dynamics in correlated materials is typically accompanied by the
interaction with bosonic particles and quasiparticles, such as phonons, plasmons, charge
density waves, photons, etc.. 
From the theoretical point of view the vastly different energy  (or time)
scales~\cite{ziman_electrons_1960,giustino_electron-phonon_2017,baroni_phonons_2001,verdozzi_classical_2006,de_melo_unified_2016,konstantinova_nonequilibrium_2018}
and the quantum nature of the involved bosonic
particles~\cite{rizzi_electron-phonon_2016,van_hest_the_role_2018,ruggenthaler_quantum-electrodynamical_2018,wang_observation_2019,karlsson_fast_2021} 
pose considerable challenges. A \emph{scalable} quantum method to model excitation and relaxation phenomena in
correlated many-body systems, reliable beyond the perturbative
regime, is crucial to simulate and interpret
experimental results and to design new materials. 
This latter aspect is especially important in view of recent 
progresses in light-enhanced phonon-induced 
superconductivity~\cite{mankowsky_nonlinear_2014,mitrano_possible_2016,sentef_theory_2016,babadi_theory_2017}, 
manipulation of thermoelectric properties with cavity 
photons~\cite{gudmundsson_time-dependent_2012,abdullah_effects_2018}, 
photonics in nanojunctions~\cite{galperin_photonics_2017}, 
exciton-phonon 
dynamics~\cite{chen_exciton-phonon_2020,stefanucci_carriers_2021,helmrich_phonon-assisted_2021}  and light-driven 
chemistry~\cite{walther_cavity_2006,hutchison_modifying_2012} to 
mention a few.  

The many-body diagrammatic theory represents a systematic way to deal with interactions
between electrons and bosonic particles.  In order to get access to the dynamical properties of the system, the
equations of motion (EOM) for the two-times electron and boson Green's functions,
hereafter referred to as the \emph{nonequilibrium Green's function} (NEGF)
theory~\cite{danielewicz_quantum_1984,van_leeuwen_introduction_2006,stefanucci_nonequilibrium_2013,karlsson_non-equilibrium_2018}, must be propagated. 
The EOMs in this case
are known as the Kadanoff-Baym equations 
(KBE)~\cite{kadanoff_quantum_1962}.  The time non-locality of the scattering
term represents the major difficulty for the full two-times propagation as it makes the
scaling at least cubic ($t_{\text{f}}^3$) with the physical propagation time
$t_{\text{f}}$~\cite{kwong_real-time_2000,dahlen_solving_2007,myohanen_many-body_2008,galperin_linear_2008,myohanen_kadanoff-baym_2009,von_friesen_successes_2009,schuler_time-dependent_2016,bittner_coupled_2018}. This hinders the possibility of resolving small energy scales as those
associated to phonons.  In the purely electronic case (no bosons)
the generalized Kadanoff-Baym ansatz
(GKBA)~\cite{lipavsky_generalized_1986} mitigates the problem of the 
cubic scaling allowing one to limit the
propagation to the time-diagonal, that is, to work with density matrices rather than with
two-times Green's functions.  One can work either with the integro-differential formulation, which has a quadratic
($t_{\text{f}}^2$) scaling in time ~\cite{hermanns_hubbard_2014, schlunzen_dynamics_2016,
bar_lev_dynamics_2014, 
latini_charge_2014,perfetto_first-principles_2015,karlsson_generalized_2018,perfetto_ultrafast_2018}, 
or with a coupled system of first-order ordinary
differential equations (ODE) thus achieving a linear ($t_{\text{f}}$) time
scaling~\cite{schlunzen_achieving_2020, joost_g1-g2_2020}.
The linear-time formulation has been already implemented to study the 
photoinduced dynamics of organic 
molecules~\cite{pavlyukh_photoinduced_2021}, carrier and exciton 
dynamics in 2D materials~\cite{perfetto_tdgw_2021} and the doublon 
production in correlated graphene clusters~\cite{Borkowski_doublon_2021}.

In our recent Letter~\cite{karlsson_fast_2021} we extended the GKBA to quantized bosonic
particles and formulated a first-order ODE, hence time-linear, scheme to treat systems with an
electron-boson ($e$-$b$) interaction. We also stated that it is possible to include the
electron-electron ($e$-$e$) interactions on equal footing.  The goal of this work is to give
an explicit demonstration of our statement.  We will further combine the GKBA+ODE
formulation with the Baym and Kadanoff
theories~\cite{baym_conservation_1961,baym_self-consistent_1962,stefanucci_nonequilibrium_2013} to generate EOM
that satisfy all fundamental conservation laws. This means that the feedback of the
electrons on the bosonic subsystem is consistently taken into account.
 Several nonperturbative methods are generated in
this way, e.g., $GW$ and $T$-matrices either without or with exchange~\cite{pavlyukh_photoinduced_2021} 
for the $e$-$e$
interaction and Ehrenfest or dressed second-order ($GD$) for the $e$-$b$ 
interaction. The
whole set of methods provides an ideal toolbox: depending on the system and the external
driving one can choose the most appropriate tool to simulate the dynamics.

The paper is organized as follows. In Section~\ref{sec:derivation} we introduce the most
general system Hamiltonian and review basic notions of the NEGF formalism. We 
then discuss diagrammatic approximations and connections between 
self-energies and high order Green's functions through the $\Phi$ 
functional of Baym. In Section ~\ref{sec:ode:gkba} we present the 
GKBA for electrons and bosons and show how to close the EOM for the electronic and
bosonic density matrices in two different ways, using either the self-energies or the
high-order Green's functions. We subsequently derive the GKBA+ODE formulation 
and discuss in  detail the diagrammatic content of the 
self-energy for all considered approximations.
Conclusions and outlook
are drawn in Section~\ref{sec:summary}.

\section{Electron-boson NEGF equations for correlated systems}%
\label{sec:derivation}                               

We consider a general electron-boson system possibly driven by 
external time-dependent fields and hence described by the Hamiltonian
\begin{equation}\label{eq:totalHamiltonian}
 \Hh(t) = \Hh_\el(t) + \Hh_\bos(t) +\Hh_\elbos(t).
\end{equation}
The electronic Hamiltonian 
\begin{align}\label{eq:H:e}
 \Hh_\el(t)&=\sum_{ij} h_{ij}(t)\hd_{i}^\dagger \hd_{j}
+\frac12\sum_{ijmn}v_{ijmn}(t)\hd_{i}^\dagger \hd_{j}^\dagger \hd_{m} \hd_{n},
\end{align}
comprises a one-body term ($h^\dagger=h$) accounting for the kinetic energy as well as the
interaction with nuclei and possible external fields and a two-body term accounting for
the Coulomb interaction between the electrons. The time-dependence of the Coulomb matrix
elements $v_{ijmn}(t)$ could be due to the adiabatic switching protocol adopted to
generate a correlated initial state. Henceforth we use latin letters to denote
one-electron states; thus $i$ is a composite index standing for an orbital degree of
freedom and a spin projection. 

The annihilation and creation operators for a bosonic mode $\bgm$, i.e., $\hat{a}_{\fbgm}$ and
$\hat{a}^{\dag}_{\fbgm}$, are arranged into a vector $(\hat{x}_{\fbgm},\hat{p}_{\fbgm})$
where $\hat{x}_{\fbgm}=(\hat{a}^{\dag}_{\fbgm}+\hat{a}_{\fbgm})/\sqrt{2}$ are the position
operators and $\hat{p}_{\fbgm}=i(\hat{a}^{\dag}_{\fbgm}-\hat{a}_{\fbgm})/\sqrt{2}$ are the
momentum operators.  The greek index $\mu=(\bgm,\xi)$ is then used to specify the bosonic
mode and the component of the vector: $\hphi_{\mu}=\hat{x}_{\fbgm}$ for $\xi=1$ and
$\hphi_{\mu}=\hat{p}_{\fbgm}$ for $\xi=2$.  We write the  bosonic Hamiltonian as
\begin{align}\label{eq:H:b}
 \Hh_\bos(t)= \sum_{\mu\nu} \Omega_{\mu\nu}(t)\hphi_\mu\hphi_\nu,
\end{align}
where $\Omega^\dagger =\Omega$ may depend on time, e.g., phonon 
drivings. 
The typical Hamiltonian for free bosons, i.e., $\Hh_\bos(t)=\sum_{\fbgm}\omega_{\fbgm}
\left(\hat{a}^{\dag}_{\fbgm}\hat{a}_{\fbgm}+\tfrac{1}{2}\right)$, 
follows from Eq.~(\ref{eq:H:b}) when setting
\begin{align}
    \Omega_{\mu\mu'}=\frac{1}{2}\delta_{\fbgm\fbgm'}\omega_{\fbgm}\delta_{\xi\xi'},
    \label{freebosOmega}
\end{align}
see also paper~II. If the bosons are photons then  
$\bgm={\bf p}$ is the momentum and $\omega_{{\bf 
p}}=c|{\bf p}|$, with $c$ the speed of light.

The electronic and bosonic subsystems interact through
\begin{align}\label{eq:H:e:b}
 \Hh_\elbos(t)&= \sum_{\mu, ij} g_{\mu, ij}(t)\hd_{i}^\dagger 
 \hd_{j}\hphi_\mu;
\end{align}
therefore electrons can be coupled to both the mode
coordinates and momenta.  Similarly to the Coulomb matrix elements 
$v$ we allow $g$ to depend on time
for possible adiabatic switchings.  

Without any loss of generality we work with an
orthonormal basis for one-electron states and one-boson states. Then the creation and
annihilation operators fulfill the standard anticommutation rules for electrons
\begin{align}
  \Big\{\hd_i,\hd^\dagger_j\Big\}&=\delta_{ij},&
  \Big\{\hd^\dagger_i,\hd^\dagger_j\Big\}&=\Big\{\hd_i,\hd_j\Big\}=0,
\end{align}
and commutation rules for bosons
\begin{equation}
  \left[\hphi_\mu,\hphi_\nu\right]=\alpha_{\mu\nu},
  \quad\quad 
  \alpha_{\mu\mu'}=-\delta_{\fbgm\fbgm'}
  \begin{pmatrix}
      0  & -i \\
      i & 0
    \end{pmatrix}_{\xi\xi'}.
\end{equation}

\subsection{NEGF formalism}

In the NEGF formalism the fundamental unknowns are the electronic lesser/greater
single-particle Green's functions
\begin{align}
  G^{<}_{ij}(t,t')&= i\langle \hat{d}_j^\dagger(t')\hat{d}_i(t)\rangle,&
  G^{>}_{ij}(t,t')&=-i\langle \hat{d}_i(t)\hat{d}_j^\dagger(t')\rangle,
  \label{eq:def:G}
\end{align}
and their bosonic counterparts
\begin{align}
  D^<_{\mu \nu}(t,t')&=D^>_{\nu\mu}(t',t) =
  -i \langle \Delta \hphi_{\nu}(t') \Delta \hphi_{\mu} (t) \rangle.\label{eq:def:D}
\end{align}
In Eq.~(\ref{eq:def:D}) we have introduced the fluctuation operators
\begin{align}
\Delta \hphi_{\nu}(t)&\equiv\hphi_{\nu}(t) -\langle \hphi_{\nu}(t)\rangle,\label{eq:def:dphi}
\end{align}
where the expectation value of the bosonic field operator $\phi_\nu(t)\equiv\langle
\hphi_{\nu}(t)\rangle$ (in contrast to the electronic case) is in general nonzero. In
Eqs.~(\ref{eq:def:G}, \ref{eq:def:D}, \ref{eq:def:dphi}) the operators are in the
Heisenberg picture and hence they depend on time.

The correlators $G^{\lessgtr}$ and $D^{\lessgtr}$ satisfy the integro-differential
Kadanoff-Baym equations (KBE) of motion. For the electronic part they read (in matrix
form):
\begin{align}
\left[ i \partial_{t} - h^{e}(t) \right] G^\lessgtr(t,t')
=\left [\Sigma^{e,\lessgtr} \cdot G^A + \Sigma^{e,R} \cdot
G^\lessgtr \right]\!(t,t'),\label{eq:e:KBE}
\end{align}
where $\left[ \mathfrak{a} \cdot \mathfrak{b}\right](t,t') \equiv 
\int d\bt\,\mathfrak{a}(t,\bt) \mathfrak{b}(\bt,t')$ is a real-time
convolution and the superscripts ``$R$'' and ``$A$'' denote the retarded and advanced
components.  The quantity $\Sigma^{e}$ is the correlation part of the electronic
self-energy; it is a functional of $G$ and $D$ through many-body diagrammatic
treatments. The time-local mean-field part is incorporated in the effective electronic
Hamiltonian $h^{e}(t)$
\begin{align}\label{eq:hele}
  h^{e}_{ij}(t)&=h_{ij}(t)+V^{{\rm HF}}_{ij}(t)+
  \sum_{\mu}g_{\mu,ij}(t)\phi_{\mu}(t),
\end{align}
where $V^{{\rm HF}}_{ij}=\sum_{mn}[v_{imnj}(t)-v_{imjn}(t)]\rho^{<}_{nm}(t)$ is the
Hartree-Fock (HF) potential written in terms of the the electronic density matrix
$\rho^<(t)$:
\begin{align}
\rho_{ij}^{\lessgtr}(t)&\equiv-i G^{\lessgtr}_{ij}(t, t),&
    [\rho^{>}_{ij}\equiv \rho^{<}_{ij}-\delta_{ij}].\label{eq:def:rho}
\end{align}

Analogously, for the bosonic propagators we have (in matrix 
form)~\cite{karlsson_non-equilibrium_2018}
\begin{align}\label{eq:b:KBE}
  \left [i \partial_t - \mh^{b}(t) \right ] \mD^\lessgtr(t,t') =
 \bga \! \left [ \mSgm^{b,\lessgtr} \cdot \mD^A +\mSgm^{b,R} \cdot \mD^\lessgtr \right]\!(t,t'),
\end{align}
where $\mSgm^{b}$ is the bosonic self-energy and 
\begin{align}
    \mh^{b}\equiv \bga(\mO+\mO^T)
    \label{hb}
\end{align}
is the
effective bosonic Hamiltonian. Like $\Sigma^{e}$ also $\mSgm^{b}$ is a functional of $G$
and $D$.  To distinguish matrices in the one-electron space from matrices in the one-boson
space we use boldface for the latters. If $\hat{H}_{\rm bos}$ is a sum
of harmonic oscillators then $\mO+\mO^T$ is proportional to the 
identity matrix in $\xi$-space, see Eq.~(\ref{freebosOmega}), and hence
$\bga(\mO+\mO^T)=(\mO+\mO^T)\bga$. For simplicity we here specialize the discussion to
this case.

To close the KBE one additionally needs to propagate the position and momentum expectation
values appearing in Eq.~\eqref{eq:hele}:
\begin{align}\label{eq:phi:eom}
  \sum\nolimits_{\nu}\left [i\delta_{\mu\nu}\partial_t - h^{b}_{\mu\nu}(t) \right ] \phi_\nu(t) 
  = \sum\nolimits_{ij}\bar{g}_{\mu,ij} \rho_{ji}(t),
\end{align}
where we have introduced 
\begin{align}
\bar{g}_{\mu.ij}\equiv 
\sum_{\nu}\alpha_{\mu\nu}g_{\nu,ij}.
\end{align}

The KBE can also be used to generate the EOM for the electronic and bosonic density
matrices.  By subtracting Eq.~\eqref{eq:e:KBE} to its adjoint and taking the equal times
limit ($t=t'$) we obtain
\begin{align}
  \frac{d}{dt}\rho^<(t)&=-i\big[h^{e}(t),\rho^<(t) \big] 
  -\left(I^{e}(t)+I^{e\,\dagger}(t)\right),
  \label{eq:eomrho:e}
\end{align}
where $I^{e}(t)$ is the right hand side of Eq.~\eqref{eq:e:KBE} calculated in $t=t'$.
Analogously, the subtraction of Eq.~\eqref{eq:b:KBE} to its adjoint and the subsequent
evaluation in $t=t'$ yields for the bosonic density matrix 
\begin{align}
\gamma_{\mu\nu}^{\lessgtr}(t)&\equiv i D^{\lessgtr}_{\mu\nu}(t, t),&
[\gamma^{>}_{\mu\nu}\equiv \gamma^{<}_{\mu\nu}+\alpha_{\mu\nu}=
\gamma^{<\,\ast}_{\mu\nu}]\label{eq:def:gamma}
\end{align}
the following equation of motion 
\begin{align}
  \frac{d}{dt}\bgg^<(t)&=-i \big[\mh^{b}(t),\bgg^<(t) \big]
  +\left(\mI^{b}(t)+\mI^{b\dagger}(t)\right),
  \label{eq:eomrho:b}
\end{align}
where $\mI^{b}(t)$ is the right hand side of Eq.~\eqref{eq:b:KBE} 
for $t=t'$. Equations~\eqref{eq:eomrho:e} and \eqref{eq:eomrho:b} are not closed because the collision
integrals $I^{e}$ and $\mI^{b}$ are still functionals of the two-times Green's functions via
the respective self-energies, $\Sigma^{e}=\Sigma^{e}[G,D;v,g]$ and
$\mSgm^{b}=\mSgm^{b}[G,D; v,g]$. In Section~\ref{sec:ode:gkba} we 
shall illustrate how to close the EOM~\eqref{eq:eomrho:e} and \eqref{eq:eomrho:b} through the GKBA for 
electrons and bosons. Preliminarly we need to develop further the 
NEGF theory and discuss diagrammatic approximations. 

\begin{figure}[tbp]
\centering  \includegraphics[width=0.9\columnwidth]{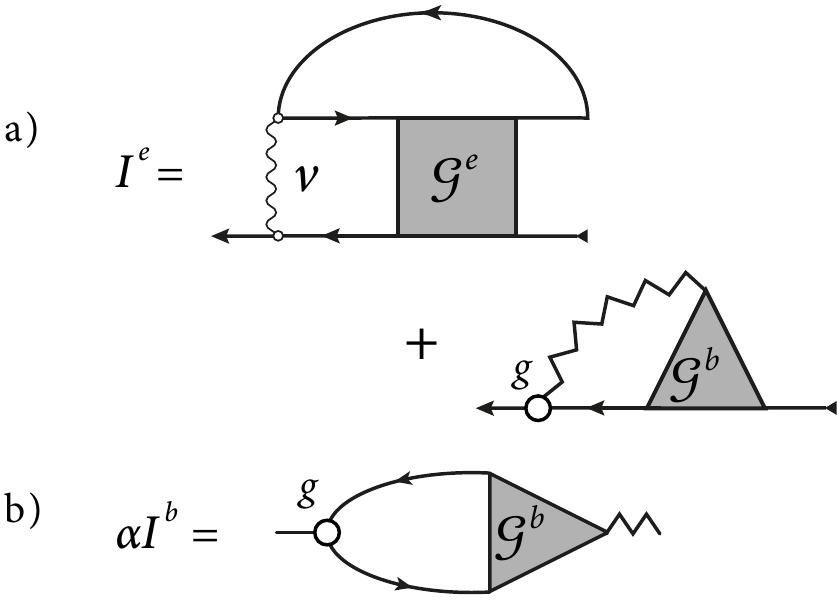}
\caption[]{Diagrammatic representation of the collision integrals in terms of high order
  Green's functions. Full lines are used for $G$, zig-zag lines are used for
  $D$, wavy lines are used for $v$ and empty circles are used for $g$.\label{fig:collisions}}
\end{figure}

We split the electronic 
self-energy into a purely electronic part 
$\Sigma^{ee}[G,v]\equiv \Sigma^{e}[G,D;v,g]_{g=0}$ and a rest, i.e., 
\begin{align}
\Sigma^{e}=\Sigma^{ee}+\Sigma^{eb}.
\end{align}
The electronic collision integral is then the sum of 
two terms, one containing $\Sigma ^{ee}$ and the other containing 
$\Sigma^{eb}$:
\begin{align}
I^{e}=I^{ee}+I^{eb}.  
\end{align}
From the first
equation of the Martin-Schwinger hierarchy for the electronic and bosonic Green's
functions one can show that the three 
collision integrals $I^{ee}$, $I^{eb}$ and $\mI^{b}$
can also be written in terms of 
two high-order Green's functions~\cite{karlsson_non-equilibrium_2018}:
\begin{align}
\cG^{e}_{imjn}(t)&=-\langle 
\hat{d}^{\dag}_{n}(t)\hat{d}^{\dag}_{j}(t)\hat{d}_{i}(t)\hat{d}_{m}(t)\rangle_{c},
\\
\cG^{b}_{\mu,ij}(t)&=\langle\hat{d}^{\dag}_{j}(t)\hat{d}_{i}(t)\hat{\phi}_{\mu}(t)\rangle_{c}.
\end{align}
The subscript ``$c$'' in the averages signifies that only the correlated part must be
retained. Like the self-energies also the high-order Green's functions are functionals
of $G$, $D$, $v$ and $g$. Pulling out from $\cG^{e}$ the  
electronic part $\cG^{ee}\equiv\cG^{e}|_{g=0}$, hence
\begin{align}
\cG^{e}=\cG^{ee}+\cG^{eb},
\label{gee+geb}
\end{align}
one finds
\begin{subequations}
  \label{eq:collisions}
\begin{align}
  I^{ee}_{lj}&=-i\sum_{imn} v_{lnmi}(t) \cG^{ee}_{imjn}(t),\label{eq:i:ee}\\
  I^{eb}_{lj}&=+i\sum_{\mu,i} g_{\mu,li}(t)\cG^{b}_{\mu,ij}(t)
  -i\sum_{imn} v_{lnmi}(t) \cG^{eb}_{imjn}(t),\label{eq:i:eb}\\
  I^{b}_{\mu\nu}&=-i\sum_{mn} \bar g_{\mu,mn}(t)\cG^{b}_{\nu,nm}(t).\label{eq:i:be}
\end{align}
\end{subequations}
Equations~(\ref{eq:collisions}) establish the relation
between the pair $\Sigma^{e}$ and $\mSgm^{b}$ and the pair $\cG^{e}$ and $\cG^{b}$; the
diagrammatic content of this relation is illustrated in Fig.~\ref{fig:collisions}.  
Notice that the diagrams for $\Sigma^{eb}(t,t')$ have either an 
$e$-$b$ vertex $g(t)$ or an $e$-$e$  vertex $v(t)$ at time $t$. 
The former provide the diagrammatic content of $\cG^{b}$ while the 
latter provide the diagrammatic content of $\cG^{eb}$, see again Eq.~(\ref{eq:i:eb}). 

It is critical to point out that for arbitrary approximations to 
$\Sigma^{eb}$ and $\mSgm^{b}$
the mixed Green's function $\cG^{b}$ entering $I^{eb}$ and $I^{b}$ need not be equal. We
therefore consider only $\Phi$-derivable
approximations~\cite{baym_self-consistent_1962,stefanucci_nonequilibrium_2013,karlsson_non-equilibrium_2018,karlsson_fast_2021} 
for these quantities.
The Baym functional $\Phi$ is expressed in terms of connected vacuum diagrams
with $e$-$b$ and $e$-$e$ vertices.
 Let  $\Phi_c[G,D;v,g]$ be the correlated part of the full Baym 
functional; $\Phi_c$ is obtained by discarding
the HF and Ehrenfest vacuum diagrams, leading to the HF potential and classical nuclear
potential appearing in Eq.~(\ref{eq:hele}). We define $\Phi^{ee}$ as 
the purely electronic part of $\Phi_c$, hence
$\Phi^{ee}[G,v]=\Phi_{c}|_{g=0}$, and write
\begin{align}
\Phi_c[G,D; v,g]= \Phi^{ee}[G,v]+\tilde{\Phi}[G,D; v,g].
\label{eq:baym}
\end{align}
The $\Phi$-derivable self-energies are then given by
(times $t$ and $t'$ on the Keldysh contour)
\begin{subequations}
\begin{align}
\Sigma^{eb}_{ij}(t,t')&=\frac{\delta\tilde{\Phi}}{\delta G_{ji}(t,t')}, 
\\
\Sigma^{b}_{\mu\nu}(t,t')&=-\frac{\delta\tilde{\Phi}}{\delta 
D_{\nu\mu}(t,t')}-\frac{\delta\tilde{\Phi}}{\delta 
D_{\mu\nu}(t',t)}.
\end{align}
\label{phisigma}
\end{subequations}
The $\Phi$-derivability guarantees that the
\emph{same} high-order Green's function $\cG^{b}$
enters $I^{eb}$ and $I^{b}$. Alternatively,  
the functional dependence
of $\cG^{b}$ on $G$ and $D$ can be directly deduced from the functional derivative of
$\tilde{\Phi}$ with respect to the $e$-$b$ coupling:
\begin{align}
  \cG^{b}_{\mu,ij}(t)&= \frac{1}{i} \,\frac{\delta \tilde{\Phi}[G,D; v,g]}{\delta 
  g_{\mu,ji}(t)}.
  \label{gbfuncder}
\end{align}

We emphasize that the $\Phi$-derivability of $\Sigma^{eb}$ and $\mSgm^{b}$
guarantees the fulfillment of all fundamental
conservation laws~\cite{nikothesis} provided that also $I^{ee}$ is calculated 
in a conserving manner. A sufficient condition for having a 
conserving $I^{ee}$ is to consider only $\Phi$-derivable electronic 
self-energies $\Sigma^{ee}_{ij}(t,t')=\delta\Phi^{ee}/\delta 
G_{ji}(t',t)$. This condition, however, is not necessary. We  
use the less stringent requirement of the symmetry of the two-particle Green's 
function (2-GF)~\cite{baym_conservation_1961}
\begin{align}
\cG^{ee}_{imjn}(t,t')\equiv
-\langle 
\hat{d}^{\dag}_{n}(t)\hat{d}^{\dag}_{j}(t')\hat{d}_{i}(t')\hat{d}_{m}(t)\rangle_{c}|_{g=0}
=\cG^{ee}_{minj}(t',t).
\label{symmcond}
\end{align}
This two-time function coincides with $\cG^{ee}(t)$ in 
Eq.~(\ref{gee+geb}) along the time diagonal, i.e., 
$\cG^{ee}(t,t)=\cG^{ee}(t)$.


\subsection{Approximations for correlated electron-boson dynamics}   

In several physical situations, e.g., phonon-induced carrier 
relaxation, Raman spectroscopy, transport through molecular junctions, etc., 
$\tilde{\Phi}$ is approximated as
\begin{align}
    \tilde{\Phi}=\raisebox{-6pt}{\includegraphics{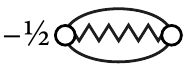}} \;\;,
    \label{Phitilde}
\end{align}
where the $e$-$b$ vertices are either bare~\cite{pellegrini_optimized_2015} (photon
fields) or statically screened~\cite{giustino_electron-phonon_2017} (phonon fields).  This
is the diagrammatic approximation that we too examine in the present work.  In most
practical implementations the boson propagator $D$ in $\tilde{\Phi}$ is frozen at its
equilibrium and noninteracting value.  We go beyond this approximation and consider
$\tilde{\Phi}$ as a functional of the fully dressed electronic {\em and} bosonic Green's
functions. Energy can then be transferred from the bosonic subsystem to the electronic
subsystem and {\em viceversa} while the total energy is conserved.  The explicit
mathematical expression of $\tilde{\Phi}$ is (time integrals are over the Keldysh contour)
\begin{equation}
\tilde{\Phi}= \frac{i}{2}\sum_{\substack{\mu\nu\\ijsq}}\int \!\! dtdt'
g_{\mu,ji}(t)D_{\mu\nu}(t,t') g_{\nu,s q}(t')
 G_{is}(t,t')   G_{q j} (t',t).
 \label{PhibGD}
\end{equation}
Through
functional derivatives with respect to $G$ and $\mD$, see 
Eqs.~(\ref{phisigma}), the chosen $\tilde{\Phi}$ leads to the
dressed second-order  self-energy ($GD$) 
for the electrons~\cite{fan_temperature_1951} and to the
bubble self-energy for the bosons~\cite{karlsson_fast_2021}, see 
Section~\ref{sec:eom} for more details.  These self-energies give the
same mixed Green's function $\cG^{b}$ from Eqs.~(\ref{eq:i:eb}) and~(\ref{eq:i:be}) since
they are derived from the same $\tilde{\Phi}$ functional.
 
\begin{figure}[]
\centering \includegraphics[width=0.9\columnwidth]{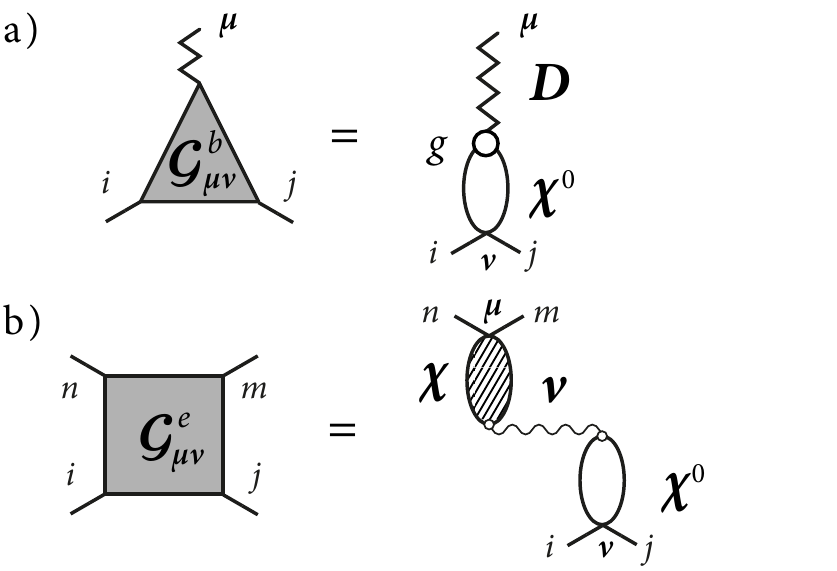}
\caption[]{(a) Mixed Green's function $\cG^{b}$ from the approximated Baym functional
in Eq.~(\ref{Phitilde}); (b) Two-particle Green's function
  $\cG^{e}=\cG^{ee}$ for $GW$, $T^{ph}$ and $T^{pp}$.\label{fig:2-GF}}
\end{figure}

We have seen that $\cG^{b}$ can also be  calculated from Eq.~(\ref{gbfuncder}).  Taking
into account that $\Phi^{ee}$ is independent of the $e$-$b$ coupling we get
\begin{align}
    \label{eq:bigG}
 \cG^{b}_{\mu,ij}(t)
 = \sum_{\nu,sq}
 \int^{t}\!\!dt' \Big\{ D^>_{\mu\nu}(t,t') g_{\nu,s q}(t')
 G^>_{is}(t,t')   G^<_{q j} (t',t)
\nonumber \\
 -(>\leftrightarrow<)\Big\}.
\end{align}
To highlight the mathematical structure in the right hand side we find useful to introduce
a composite index for pairs of electron indices. Without any risk of ambiguity we use
greek letters also for such composite index:
\begin{align}
  g_{\mu,ij}&= g_{\mu,\begin{subarray}{c}j\\i\end{subarray}}=g_{\mu\nu},&
   \cG^{b}_{\mu,ij}&\rightarrow 
   \cG^{b}_{\mu,\begin{subarray}{c}j\\i\end{subarray}}=\cG^{b}_{\mu\nu},&\nu=
   \begin{pmatrix}j\\i\end{pmatrix}.
\end{align}
This mathematical notation expresses the physical notion that we need 
two fermions to make a boson.
We can then rewrite Eq.~(\ref{eq:bigG}) in a compact matrix form as
\begin{align}
\mcG^{b}(t)&=i\int^{t}\!\!dt'\Big\{
 \mD^{>}(t,t')\bcallg(t')\mchi^{0,<}(t',t)
 -(>\leftrightarrow<)\Big\},\label{eq:G2:b:0}
\end{align}
where, consistently with our notation, the matrices with greek indices are represented by
boldface letters. In Eq.~(\ref{eq:G2:b:0}) we have defined the noninteracting response
function
\begin{align}
\chi^{0,\lessgtr}_{\mu\nu}(t',t)=
\chi^{0,\lessgtr}_{\begin{subarray}{c}qj\\si\end{subarray}}(t',t)\equiv
-iG^\lessgtr_{q j} (t',t)G^\gtrless_{is}(t,t').
\label{chi0eb}
\end{align}
The diagrammatic representation of Eq.~(\ref{eq:G2:b:0}) is given in 
Fig.~\ref{fig:2-GF}(a).

Let us now discuss the 2-GF $\cG^{e}$. The functional $\tilde{\Phi}$ in Eq.~(\ref{Phitilde}) 
is independent of the Coulomb interaction $v$ and therefore 
$\cG^{eb}=0$, or equivalently $\cG^{e}=\cG^{ee}$, see 
Eq.~(\ref{gee+geb}). For $\cG^{ee}$ we consider a large class of perturbative and
nonperturbative approximations like the second-Born (2B), $GW$ and $T$-matrix in the
particle-hole ($T^{ph}$) and particle-particle ($T^{pp}$) channels as well as $GW$ plus
exchange ($X$), $T^{ph}+X$ and $T^{pp}+X$~\cite{pavlyukh_photoinduced_2021}. 
In Ref.~\cite{pavlyukh_photoinduced_2021} we have also shown how to 
include three-particle correlations 
through a generalization of the GKBA to the three-particle Green's 
function. The method
has been dubbed the Faddeev approach and it is particularly suited to study the correlated
electron dynamics in molecules after an inner-valence ionization.
All the aforementioned approximations to $\cG^{ee}$ satisfy the 
symmetry condition in Eq.~(\ref{symmcond});
hence the resulting theory is fully conserving.

For the more familiar  $GW$, $T^{ph}$ and $T^{pp}$ approximations  
the 2-GF satisfies an RPA-like equation whose diagrammatic 
representation can be found in, e.g., Figure 1 of 
Ref.~\cite{pavlyukh_photoinduced_2021}. Therefore, the 2-GF
can be written as (in matrix 
form)
\begin{align}
\mcG^{ee}(t)&=-i\int^{t}\!\!dt'\Big\{ \mchi^{>}(t,t')\mv(t')\mchi^{0,<}(t',t)
-(>\leftrightarrow<)\Big\},\label{eq:G2:e:0}
\end{align}
where 
\begin{align}
\mchi^{\lessgtr}=
(\bgd+\mchi^R\cdot \mv)\cdot\mchi^{0,\lessgtr}\cdot(\mv\cdot\mchi^A+\bgd)
\label{chilessgtr}
\end{align}
and
\begin{subequations}
  \label{eq:rpa}
\begin{align}
  \mchi^{R/A}&=\mchi^{0,R/A}+\mchi^{0,R/A}\cdot \mv\cdot\mchi^{R/A},\label{eq:rpa:1}\\
  &=\mchi^{0,R/A}+\mchi^{R/A}\cdot\mv\cdot\mchi^{0,R/A}.\label{eq:rpa:2}
\end{align}
\end{subequations}
In Eq.~(\ref{chilessgtr}) the quantity $\bgd$ stands for the Dirac delta in time and the
Kroenecher delta in the two-electron space, hence for any two-time correlator $\bcallC$ we
have $[\bgd\cdot \bcallC](t,t')=\int
d\bar{t}\,\bgd(t-\bar{t})\bcallC(\bar{t},t')=\bcallC(t,t')$.  Depending on the
approximation the matrices in the two-electron space $\mcG^{ee}$, $\mchi^{0}$ and $\mv$ are
defined in Table~\ref{tab:1}. The perturbative 2B approximation is simply obtained by
replacing the dressed $\mchi$ with $\mchi^{0}$ in Eq.~(\ref{eq:G2:e:0}).  Notice that the
matrix $\mchi^{0}$ in Eq.~(\ref{chi0eb}) coincides with the matrix $\mchi^{0}$ in
Eq.~(\ref{eq:G2:e:0}) only for the 2B and $GW$ approximations. To keep the notation as
light as possible we however use the same symbol even when they are different ($T^{ph}$
and $T^{pp}$ approximation) and make the reader notice the slight abuse of notation when
this occurs. The diagrammatic representation of Eq.~(\ref{eq:G2:e:0}) is given in
Fig.~\ref{fig:2-GF}(b). 

In the next section we derive the EOM for $\rho^{<}$, $\bgg^{<}$,
$\mcG^{b}$ and $\mcG^{ee}$  in the $GD$ approximation and in
all approximations of Table~\ref{tab:1}. We also show how to modify the EOM
when exchange is included and provide the diagrammatic content of the corresponding
self-energies.

\begin{table}[t!]
  \caption{\label{tab:1} Definitions of electronic two-particle tensors. The vertically
    grouped indices are combined into one (greek) super-index.}  \renewcommand{\arraystretch}{1.4}
  \begin{ruledtabular}
    \begin{tabular}{LLLL}
      \text{Quantity}& \text{2B and } GW & \multicolumn{1}{C}{T^{pp}} & \multicolumn{1}{C}{T^{ph}}\\\hline
      i\mchi_{\begin{subarray}{c}13\\24\end{subarray}}^{0,\lessgtr}(t,t')&
      G^{\lessgtr}_{13}(t,t')G^{\gtrless}_{42}(t',t)&
      -G^{\lessgtr}_{13}(t,t')G^{\lessgtr}_{24}(t,t')&
      -G_{13}^{\lessgtr}(t,t')G^{\gtrless}_{42}(t',t)\\  [1pt]
      \mcG^{ee}_{\begin{subarray}{c}13\\24\end{subarray}}&
      \mathcal{G}^{ee}_{4132}&
      \mathcal{G}^{ee}_{1234}&
      \mathcal{G}^{ee}_{1432}\\[1pt] 
      \mh^{e}_{\begin{subarray}{c}13\\24\end{subarray}}&
      h^{e}_{13}\delta_{42}-\delta_{13}h^{e}_{42}&
      h^{e}_{13}\delta_{24}+\delta_{13}h^{e}_{24}&
      h^{e}_{13}\delta_{42}-\delta_{13}h^{e}_{42}\\ 
      \mv_{\begin{subarray}{c}13\\24\end{subarray}} &
      v_{1432}&
      v_{1243}&
      v_{1423}\\[1pt]
      \mrho^{<}_{\begin{subarray}{c}13\\24\end{subarray}}&
      \rho^<_{13}\rho^>_{42}&
      \rho^<_{13}\rho^<_{24}&
      \rho^<_{13}\rho^>_{42}\\
       \end{tabular}
    \end{ruledtabular}
\end{table}

\section{GKBA+ODE scheme for  NEGF simulations \label{sec:ode:gkba}}

\subsection{GKBA for electrons and bosons}  
\label{sec:EBGKBA}
  
A way to close the EOM for the density matrices consists in 
implementing 
the GKBA for electrons~\cite{lipavsky_generalized_1986} and our 
recently proposed GKBA for bosons~\cite{karlsson_fast_2021} 
\begin{align}
  G^{\lessgtr}(t,t')&=-G^{R}(t,t')\rho^{\lessgtr}(t')+\rho^{\lessgtr}(t)G^{A}(t,t'),\label{eq:e:gkba}\\
  \mD^{\lessgtr}(t,t')&=\mD^{R}(t,t')\bga\bgg^{\lessgtr}(t')-\bgg^{\lessgtr}(t)\bga \mD^{A}(t,t'),\label{eq:b:gkba}
\end{align}
combined with the mean-field form of the retarded propagators:
\begin{align}
G^{R}(t,t')&=-i\theta(t-t')T\left\{e^{-i \int_{t'}^t d\tau\, h^{e}(\tau)}\right\},\label{eq:gr:hf}\\
\mD^{R}(t,t')&=-i\bga \theta(t-t') T\left\{e^{-i \int_{t'}^t d\tau\, 
\mh^{b}(\tau)}\right\}.\label{eq:dr:hf}
\end{align}
We mention that more advanced propagators can be used without affecting the scaling of the
numerical
solution~\cite{marini_ab_2008,marini_dynamical_2003,haug_interband_1992,bonitz_non-lorentzian_1999,latini_charge_2014}.
Once the GKBA is applied to a given approximation to the self-energies (or equivalently
high-order Green's functions) both $I^{e}$ and $\mI^{b}$ become functionals of $\rho^<$
and $\bgg^{<}$; hence Eqs.~(\ref{eq:phi:eom}), (\ref{eq:eomrho:e}) and (\ref{eq:eomrho:b})
become a closed system of integro-differential equations for the one-time unknown
functions $\phi(t)$, $\rho^{<}(t)$ and $\bgg^{<}(t)$. We refer to this approach as the
GKBA+KBE.

For purely electronic systems an efficient implementation of the GKBA equation of motion
(\ref{eq:eomrho:e}) has been recently
proposed~\cite{schlunzen_achieving_2020,joost_g1-g2_2020}. The main feature is the linear
scaling with the maximum propagation time for the 2B, $GW$ and $T$-matrix approximations.
In Ref.~\cite{pavlyukh_photoinduced_2021} the class of approximations has been further
extended to include exchange effects and even three-particle correlations. The question
what is the most general approximation to $\tilde{\Phi}$ for preserving the time-linear
scaling property is still open.  In this work we make a step in this direction and show
that the approximate $\tilde{\Phi}=$\raisebox{-6pt}{\includegraphics{Fig2}} (discussed in the previous
section) does not affect the overall time scaling.

\subsection{GKBA form of ${\mbox{\boldmath $\mathcal{G}$}}^{b}$ and
  ${\mbox{\boldmath $\mathcal{G}$}}^{e}$}

The Green's functions $\mcG^{b}$ and $\mcG^{e}=\mcG^{ee}$ are prerequisites for reformulating the
GKBA+KBE equations in terms of first-order ODE, thus achieving a linear time-scaling scheme.  The
purpose of this section is to implement the GKBA and transform these 
high-order Green's
functions into functionals of $\rho^{<}$ and $\bgg^{<}$.

Let us consider first $\mcG^{b}$.  Evaluating the noninteracting response functions of
Eq.~(\ref{chi0eb}) with the GKBA in Eq.~(\ref{eq:e:gkba}) we find a 
sum of products between matrices in the two-electron space
\be
\mchi^{0,\lessgtr}(t,t')=\mP^R(t,t')\mrho^{\lessgtr}(t')-\mrho^{\lessgtr}(t)\mP^A(t,t'),
\label{eq:gkba:chi0}
\ee
where $\mP^{A}(t,t')=\left[\mP^{R}(t',t)\right]^\dagger$ 
and for $t>t'$ the particle-hole
propagator fulfills the equation of motion
\begin{align}
  i\frac{d}{dt}\mP^R(t,t')&=\mh^{e}(t)\mP^R(t,t'),
\label{eq:EOM:P}
\end{align}
with boundary condition $i\mP^R(t^{+},t)=-\mathbb{1}$ and $\mP^R(t,t')=0$ for $t<t'$. The
boldface quantities $\mrho^{<}(t)$ and $\mh^{e}(t)$ are matrices in the two-electron space and they
are given in Table~\ref{tab:1} under the column 2B and $GW$.  The matrix $\mrho^{>}$ is
obtained from $\mrho^{<}$ by changing the one-electron matrices
$\rho^{\lessgtr}\to\rho^{\gtrless}$.  Substituting
Eqs.~(\ref{eq:gkba:chi0}) and (\ref{eq:b:gkba}) into 
Eq.~\eqref{eq:G2:b:0} we find
\begin{align}
  \mcG^{b}(t)&=-i\int^{t}\!\!dt'\mD^R(t,t')\malpha\mPsi^{b}(t')\mP^A(t',t),
\label{eq:gb:explicit}
\end{align}
with
\begin{align}
  \mPsi^{b}(t)\equiv \bgg^{>}(t)\bcallg(t)\mrho^{<}(t)-\bgg^{<}(t)\bcallg(t) \mrho^{>}(t).
  \label{Psi:b:def}
\end{align}
Equation (\ref{eq:gb:explicit}) is a functional of the bosonic and fermionic density
matrices through the definitions in Table~\ref{tab:1}, Eq.~(\ref{eq:dr:hf}) and
Eq.~(\ref{eq:EOM:P}).

Next we consider the 2-GF $\mcG^{ee}$. Using the GKBA to evaluate the noninteracting
response function $\mchi^{0}$ in one of the approximations of Table~\ref{tab:1} we always
find Eq.~(\ref{eq:gkba:chi0}) where $\mP^{R}$ fulfills the same equation of motion as in
Eq.~(\ref{eq:EOM:P}) but with boundary conditions $i\mP^R(t^{+},t)=-\mathbb{1}$ for 2B and
$GW$ and $i\mP^R(t^{+},t)=\mathbb{1}$ for the $T$-matrix approximations. Another
important difference is that the definition of the matrices $\mrho^{\lessgtr}$ and
$\mh^{e}(t)$ changes by changing approximation according to Table~\ref{tab:1}. Using this
result the retarded and advanced noninteracting response function, 
i.e., 
\begin{align}
    \mchi^{0,R/A}(t,t')\equiv \pm\theta(\pm t\mp
t')[\mchi^{0,>}(t,t')-\mchi^{0,<}(t,t')],
\end{align}
read
\begin{subequations}
\begin{align}
\mchi^{0,R}(t,t')&=\mP^{\rm R}(t,t')\mrho^{\Delta}(t'),
\\
\mchi^{0, A}(t,t')&=\mrho^{\Delta}(t)\mP^{\rm A}(t,t').
\label{RPAchiRA}
\end{align}
\label{chi0ra}
\end{subequations}
where
\begin{align}
  \mrho^\Delta(t)&\equiv \mrho^{>}(t)-\mrho^{<}(t).
  \label{rhodelta}
\end{align} 

In Ref.~\cite{pavlyukh_photoinduced_2021} we have shown that inserting Eqs.~(\ref{chi0ra})
into Eq.~(\ref{eq:rpa}) for $\mchi^{R/A}$ and then using Eq.~(\ref{eq:gkba:chi0}) for
$\mchi^{0,\lessgtr}$, the lesser/greater interacting response function in
Eq.~(\ref{chilessgtr}) can be written as
\begin{align}
  \mchi^{\lessgtr}
 =\mPi^R\mrho^{\lessgtr}\cdot\big(\bgd+\mv\mrho^\Delta\mPi^A\big)-
  \big(\bgd+\mPi^R\mrho^\Delta\mv\big)\cdot\mrho^{\lessgtr}\mPi^A,
  \label{eq:chi:gtrless}
\end{align}
where the dressed propagator $\mPi^{R}(t,t')=\left[\mPi^{A}(t',t)\right]^\dagger$ fulfills
the RPA equation
\begin{subequations}
\begin{align}
  \mPi^{R}-\mP^{R}&=\mPi^{R}\cdot\mrho^\Delta\mv\mP^{R}
  =\mP^{R}\mrho^\Delta\mv\cdot\mPi^{R},\\
  \mPi^{A}-\mP^{A}&=\mPi^{A}\cdot\mv\mrho^\Delta\mP^{A}
  =\mP^{A}\mv\mrho^\Delta\cdot\mPi^{A}.
\end{align}
\label{eq:rpa:PiR}
\end{subequations}

\noindent
For later purposes we also observe that taking into account the equation of motion
(\ref{eq:EOM:P}) for $\mP^{R}$ together with its boundary condition, the dressed
propagators satisfy a simple EOM
\begin{subequations}
\begin{align}
  i\frac{d}{dt}\mPi^{R}(t,t')&=
  \big[\mh^{e}(t)+a \,\mrho^{\Delta}(t)\mv (t)\big]\mPi^{R}(t,t'),
  \\
    -i\frac{d}{dt}\mPi^{A}(t',t)&=
  \mPi^{A}(t',t)\big[\mh^{e}(t)+a \,\mv(t) \mrho^{\Delta}(t)\big],
\end{align}
\label{eq:Pi:EOM}
\end{subequations}
\noindent
where the constant $a$ depends on the approximation:
$a=0$ in 2B, $a=-1$ in $GW$, $a=1$ in $T^{ph},\; T^{pp}$.
The two-time function $\mPi^{R}$ can be interpreted 
as a dressed particle-hole (for $GW$ and $T^{ph}$) or 
particle-particle (for $T^{pp}$) propagator.

We have now all the ingredients to obtain the GKBA form of the 2-GF, 
and hence to transform $\mcG^{ee}$ into a functional of $\rho^{<}$.
We substitute Eq.~(\ref{eq:chi:gtrless}) for $\bgx^{\lessgtr}$ and
Eq.~(\ref{eq:gkba:chi0}) for $\bgx^{0,\lessgtr}$ into Eq.~(\ref{eq:G2:e:0}) and find
\begin{widetext}
\begin{align}
\mcG^{ee}(t)&=i
\Big[\Big(\underbrace{\mPi^{R}\mrho^{>}\cdot(\bgd+\mv\,\mrho^{\Delta}\mPi^{A})
-(\bgd+\mPi^{R}\mrho^{\Delta}\,\mv)\cdot\mrho^{>}\mPi^{\rm 
A}}_{\bgx^{>}}\Big)\cdot \mv\underbrace{\mrho^{<} 
\mP^{A}}_{-\bgx^{0,<}}\Big](t,t)
-\Big[>\;\leftrightarrow\; <\Big](t,t)
\nn\\
&=i\Big[\mPi^{R}\mrho^{>} \mv\mrho^{<}\cdot \mP^{A}
+\mPi^{R}
(\mrho^{>}\mv\mrho^{\Delta}-\mrho^{\Delta}\mv\mrho^{>})\cdot
\mPi^{\rm A}\cdot\mv\mrho^{<} \mP^{ A}
\Big](t,t)
\nonumber \\
&-i\Big[\mPi^{R}\mrho^{<} \mv\mrho^{>}\cdot \mP^{ A}
+\mPi^{R}
(\mrho^{<}\mv\mrho^{\Delta}-\mrho^{\Delta}\mv\mrho^{<})\cdot
\mPi^{ A}\cdot\mv\mrho^{>} \mP^{A}
\Big](t,t)
\nn\\
&=i\Big[\mPi^{R}\,(\mrho^{>}\mv\mrho^{<}-\mrho^{<}\mv\mrho^{>})\cdot \mP^{ A}+
\mPi^{R}\,(\mrho^{>}\mv\mrho^{<}-\mrho^{<}\mv\mrho^{>})
\cdot \mPi^{A}\cdot \mv\,\mrho^{\Delta}\mP^{A}\Big](t,t)
\nn\\
&=i\Big[\mPi^{R}\,(\mrho^{>}\mv\mrho^{<}-\mrho^{<}\mv\mrho^{>})\cdot \mPi^{A}\Big](t,t),
\label{callGGWXgkba}
\end{align}
\end{widetext}
where in the second equality we have observed that $[\mPi^{A}\cdot \mv\,\mrho^{\lessgtr}
  \mP^{A}](t,t)=0$ since $\mPi^{A}(t,\bar{t})$ contains a $\theta(\bar{t}-t)$ and
$\mP^{A}(\bar{t},t)$ contains a $\theta(t-\bar{t})$.  Making explicit the time integration
we recognize the same mathematical structure of the mixed Green's function $\mcG^{b}$ in
Eq.~(\ref{eq:gb:explicit})
\begin{align}
\mcG^{ee}(t)=i\int^{t}\!dt'\;
\mPi^{R}(t,t')\,\mPsi^{e}(t') \mPi^{A}(t',t),
\label{GeGKBA}
\end{align}
where we have defined
\begin{align}
  \mPsi^{e}(t)&\equiv \mrho^{>}(t)\mv(t)\mrho^{<}(t)-\mrho^{<}(t)\mv(t) \mrho^{>}(t).
  \label{psie}
\end{align} 
Equation (\ref{GeGKBA}) is a functional of the bosonic and fermionic density
matrices through the definitions in Table~\ref{tab:1} and 
Eqs.~(\ref{eq:Pi:EOM}).

\subsection{EOM for ${\mbox{\boldmath $\mathcal{G}$}}^{b}$ and ${\mbox{\boldmath $\mathcal{G}$}}^{e}$ \label{sec:eom}}

Differentiating Eq.~\eqref{eq:gb:explicit} with respect to $t$ and taking into account
that $\mD^R(t,t')$ defined in Eq.~(\ref{eq:dr:hf}) satisfies for $t>t'$ the equation
$i\frac{d}{dt}\mD^R(t,t')=\mh^{b}(t)\mD^R(t,t')$ we find
\begin{align}
  i\frac{d}{dt}\mcG^{b}(t)&=-\mPsi^{b}(t)+\mh^{b}(t)\mcG^{b}(t)-\mcG^{b}(t)\mh^{e}(t),
\label{eq:EOM:gb}
\end{align}
where we also used the equation of motion (\ref{eq:EOM:P}) for
$\mP^{A}(t',t)=[\mP^{R}(t,t')]^{\dag}$. 
We notice that Eq.~(\ref{eq:EOM:gb}) differs from the EOM in 
Ref.~\cite{karlsson_fast_2021} for 
the minus sign in front of $\mPsi^{b}$. This is due to 
the fact that a minus sign has been introduced in the present definition 
of $\rho^{>}$, see Eq.~(\ref{eq:def:rho}).

Similarly, differentiating Eq.~(\ref{GeGKBA})
with respect to $t$ and taking into account the equation of motion (\ref{eq:Pi:EOM}) for
$\mPi^{R/A}$ we find
\begin{multline}
  i \frac{d}{dt}\mcG^{ee}(t)=-\mPsi^{e}(t)+\left[\mh^{e}(t) 
   +a\mrho^\Delta(t)\mv(t)\right]\mcG^{ee}(t)\\
  -\mcG^{ee}(t) \left[\mh^{e}(t) +a\mv(t) 
    \mrho^\Delta(t)\right].
  \label{eq:G2:X:init}
\end{multline}
Equations (\ref{eq:EOM:gb}) and (\ref{eq:G2:X:init}) together with 
the equation of motion for $\phi(t)$ [Eq.~(\ref{eq:phi:eom})] and the equations of motion
for the electronic and bosonic density matrices 
[Eqs.~(\ref{eq:eomrho:e}) and
  (\ref{eq:eomrho:b})],
form a closed system of first-order ODE to study the dynamics of interacting electrons and
bosons in a large class of approximations, see also below. This is the GKBA+ODE scheme. 

\begin{figure}[]
  \centering\includegraphics[width=0.95\columnwidth]{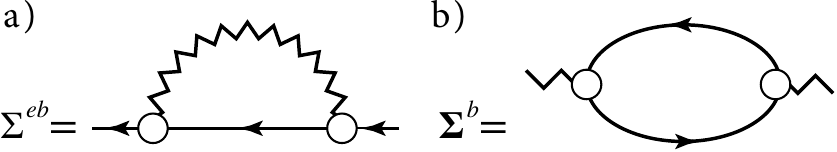}  
\caption[]{Bosonic self-energy for electrons (a) and electronic self-energy for
  bosons (b).\label{fig:sgm:eb:b}}
\end{figure}

\begin{figure*}[tbp]
  \centering\includegraphics[width=0.95\textwidth]{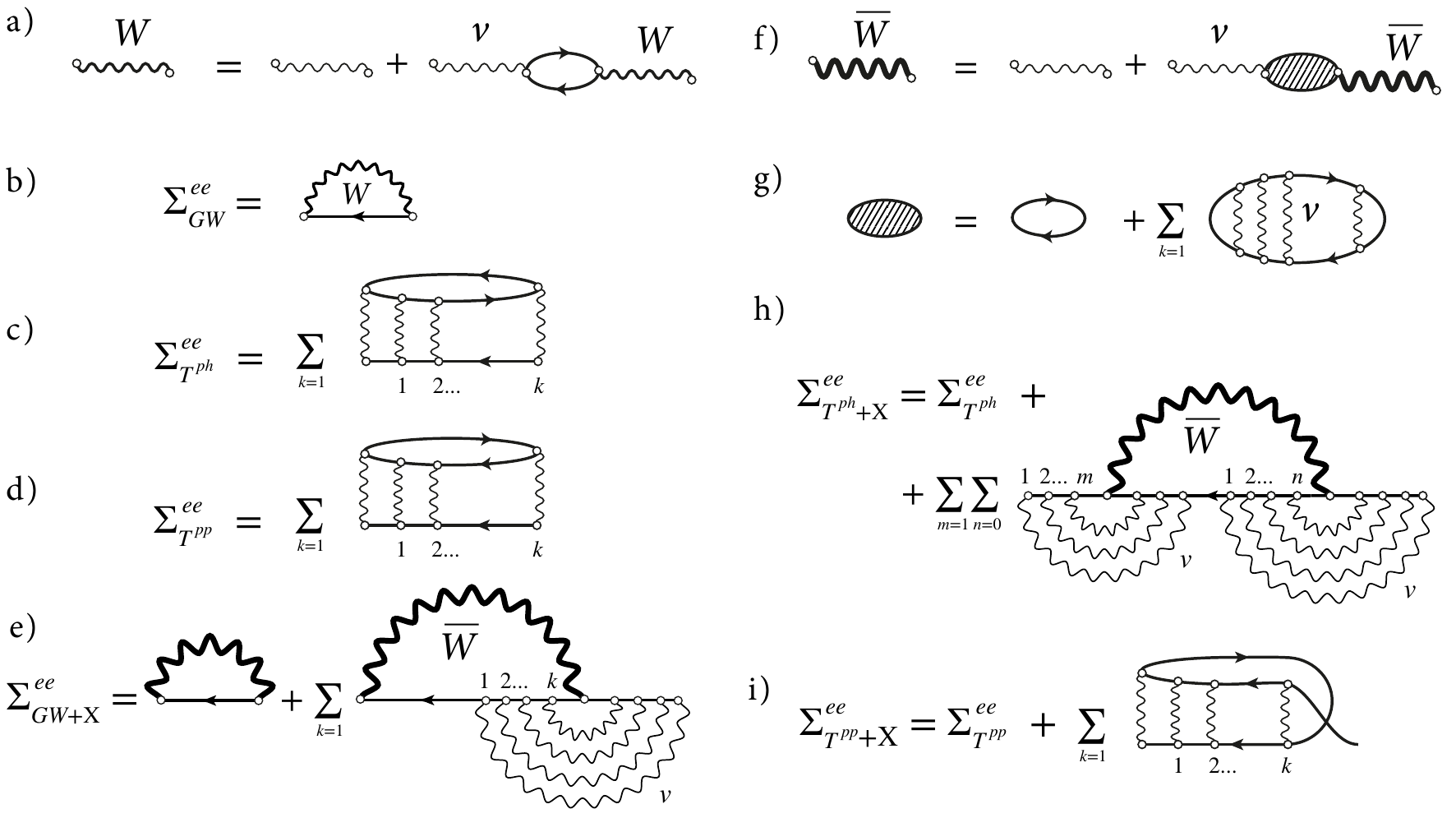}  
\caption[]{All possible self-energies $\Sigma^{ee}$ that can be 
implemented in the GKBA+ODE scheme. (a) RPA screened interaction; (b) 
$GW$ self-energy with RPA screened interaction; (c) $T$-matrix self-energy in the particle-hole 
channel; (d) $T$-matrix self-energy in the particle-particle 
channel; (e) $GW+X$ approximation consisting of a $GW$-like diagram 
and an infinite series of exchange diagrams. The screened interaction 
is $\bar{W}$ defined in (f) where the polarization (g) differs from the 
bare one as it accounts for multiples electron-hole scatterings. (h) 
$T^{ph}+X$ approximation consisting of the standard $T^{ph}$ diagrams 
and an infinite series of exchange diagrams -- notice the appearance of 
$\bar{W}$. (i) $T^{pp}+X$ approximation.
\label{fig:self-energy}}
\end{figure*}

The GKBA+ODE scheme is equivalent to the original GKBA+KBE integro-differential equations
with electronic self-energy $\Sigma^{e}=\Sigma^{ee}+\Sigma^{eb}$ and bosonic self-energy $\mSgm^{b}$. For the
approximate functional $\tilde{\Phi}$ in Eq.~(\ref{PhibGD}) we have
that $\Sigma^{eb}$ is the 
 $GD$ self-energy calculated using {\em dressed} Green's functions $G$ and $\mD$, see
Fig.~\ref{fig:sgm:eb:b}(a).  The dressed Green's function $\mD$ differs from its
equilibrium counterpart since bosons receive an electronic feedback through
$\mSgm^{b}$. The latter is in turn evaluated in accordance with the $\Phi$-derivability
theory of Baym and it is therefore given by the electronic bubble, see
Fig.~\ref{fig:sgm:eb:b}(b).  The electronic self-energy due to the 
$e$-$e$
interaction $\Sigma^{ee}$ can instead be approximated in several ways.  In addition to the
perturbative 2B approximation [which corresponds to set $a=0$ in Eq.~(\ref{eq:G2:X:init})]
the two-electron index-order outlined in Table~\ref{tab:1} is equivalent to the
implementation of the $GW$ approximation ($a=-1$), see Fig.~\ref{fig:self-energy}(a,\,b) or the
$T$-matrix approximation ($a=1$) in the particle-hole and particle-particle channels, see
Fig.~\ref{fig:self-energy}(c,d).

The 2-GF in the $GW$ and $T$-matrix approximations solve the Bethe-Salpeter 
equation (BSE) with kernel
$\mv$ given in Table~\ref{tab:1} under the respective ($GW$ or $T^{ph}$) column.  
Exchange
effects to all orders can be included using the kernel $\mw$ (same 
index-order convention as
$\mv$) where 
\begin{align}
    w_{imnj}=v_{imnj}-v_{imjn},
\end{align}
i.e., the sum of the direct and exchange ($X$)
Coulomb integrals. The addition of exchange to the BSE kernel 
preserves the symmetry condition in Eq.~(\ref{symmcond}) and, 
therefore, these approximations too are conserving. 
The GKBA form of the resulting $\mcG^{ee}$ satisfies Eq.~(\ref{eq:G2:X:init}) where
all $\mv$'s are replaced by $\mw$ -- hence also the definition of $\mPsi^{e}$ in Eq.~(\ref{psie})
changes. We refer to these approximations as the $GW+X$ method if $\mw$ is written
with the $GW$ index-order and the $T^{ph}+X$ method if $\mw$ is written in the $T^{ph}$
index-order~\cite{pavlyukh_photoinduced_2021}.  As there is a one-to-one diagrammatic correspondence 
between $\cG^{ee}$ and
$\Sigma^{ee}$ it is instructive to work out the self-energy diagrams in these
two approximations. We anticipate that $\Sigma^{ee}$ is not 
$\Phi$-derivable in $GW+X$ and $T^{ph}+X$ (nonetheless the theory 
is conserving).

In Fig.~\ref{fig:self-energy}(e) we show the $GW+X$ self-energy. It
consists of a $GW$-like diagram and of an infinite series of exchange diagrams. The
screened interaction $\bar{W}$ differs from the RPA $W$ in 
Fig.~\ref{fig:self-energy}(a) since the polarization contains a
nonperturbative vertex correction describing the multiple scattering 
between an electron-hole pair, see Fig.~\ref{fig:self-energy}(f,\,g). 
If we replace $\bar{W}$ with the RPA $W$ 
and restrict the sum over 
$k$ to $k=1$ then the $GW+X$ self-energy reduces to the SOSEX self-energy 
introduced in Ref.~\cite{ren_beyond_2015}, see also 
Refs.~\cite{maggio_gw_2017,pavlyukh_dynamically_2020,wang_assessing_2021}.

The $T^{ph}+X$ self-energy is illustrated in Fig.~\ref{fig:self-energy}(h). In addition to
the standard $T^{ph}$ diagrams, see again Fig.~\ref{fig:self-energy}(d), this
approximation contains a $GW$-like diagram decorated by an infinite series of exchange
terms at both vertices. It is worth noticing that the sum over the interactions start from
unity on the left and from zero on the right, hence no $GW$-like 
diagram is contained in here. Interestingly, the screened interaction is
the same as in the $GW+X$ approximations, i.e., it is the  
$\bar{W}$ of Fig.~\ref{fig:self-energy}(f). We finally observe that
the Bethe-Salpeter equation with kernel $\mw$ in the particle-particle channel would lead to a
multiple counting of the same diagram since exchanging two particles 
twice is equivalent to no exchange.
Therefore the proper way of constructing the $T^{pp}+X$ approximation is
depicted in Fig.~\ref{fig:self-energy}(i), where all internal lines 
are bare $e$-$e$ interactions. The
EOM for $\mcG^{ee}$ in $T^{pp}+X$ is the same as in the $T^{pp}$ approximation. The
difference appears in the EOM for $\rho^{<}$; the collision integral $I^{ee}$ of
Eq.~(\ref{eq:i:ee}) should in this case be calculated with $v_{imnj}\to w_{imnj}$.

As Fig.~\ref{fig:self-energy} shows, the GKBA+ODE scheme can be implemented for a large
number of  methods. For all of them the EOM 
have the same mathematical structure:
\begin{widetext}
\begin{subequations}
\begin{align}
i\frac{d}{dt}\phi_\mu(t)&=  h^{b}_{\mu\nu}(t)\phi_\nu(t) 
  +\sum\nolimits_{ij}\bar{g}_{\mu,ij} \rho_{ji}(t),
\label{GKBA-ODE1}
\\
i\frac{d}{dt}\rho^<_{lj}(t)&=
\big[h^{e}(t),\rho^<(t) \big] _{lj}
  +\left(-c\sum_{imn} \big(v_{lnmi}(t)-x'\,v_{lnim}(t)\big) 
  \cG^{ee}_{imjn}(t)+d \sum_{\mu,i} 
  g_{\mu,li}(t)\,\cG^{b}_{\mu,ij}(t)-(l\leftrightarrow j)^{\ast}\right),
\label{GKBA-ODE2}
\\
i\frac{d}{dt}\gamma_{\mu\nu}^<(t)&= 
\big[\mh^{b}(t),\bgg^<(t) \big]_{\mu\nu}
  +\left(d\sum_{mn} \bar 
  g_{\mu,mn}(t)\,\cG^{b}_{\nu,nm}(t)-(\mu\leftrightarrow\nu)^{\ast}\right),
\label{GKBA-ODE3}
\\
i\frac{d}{dt}\mcG^{b}(t)&=-\mPsi^{b}(t)+\mh^{b}(t)\,\mcG^{b}(t)-\mcG^{b}(t)\,\mh^{e}(t),
\label{GKBA-ODE4}
\\
i\frac{d}{dt}\mcG^{ee}(t)&=-\mPsi^{e}(t)+\left[\mh^{e}(t)  
+a\mrho^\Delta(t)\,\mathbfit{u}_{x}(t)\right]\mcG^{ee}(t)
  -\mcG^{ee}(t) \left[\mh^{e}(t) +a\,\mathbfit{u}_{x}(t) \,
  \mrho^\Delta(t)\right],
  \label{GKBA-ODE5}
\end{align}
\label{GKBA-ODE}
\end{subequations}
\end{widetext}
where the control parameters $a$, $c$, $x$, $x'$ and $d$  allows to 
switch between different methods
\begin{flalign*}
&\begin{aligned}
&a=\left\{\begin{array}{rl}
{\rm any} & \quad\quad {\rm HF} \\
0 & \quad\quad{\rm 2B} \\
-1 & \quad\quad GW+(X) \\
1 & \quad\quad T^{ph}+(X),\;T^{pp}+(X)
\end{array}\right.
\end{aligned}&&
\end{flalign*}
\begin{flalign*}
&\begin{aligned}
c=\left\{\begin{array}{rl}
0 & \quad\quad\;\;\;{\rm HF} \\
1 & \quad\quad\;\;\;{\rm otherwise}
\end{array}\right.
\end{aligned}&&
\end{flalign*}
\begin{flalign*}
&\begin{aligned}
x=\left\{\begin{array}{rl}
0 & \quad\quad\;\;\; GW,\;T^{ph},\;T^{pp}+(X) \\
1 & \quad\quad\;\;\;  {\rm 2B},\;GW+X,\;T^{ph}+X
\end{array}\right.
\end{aligned}&&
\end{flalign*}
\begin{flalign*}
&\begin{aligned}
x'=\left\{\begin{array}{rl}
1 & \quad\quad\;\;\; T^{pp}+X \\
0 & \quad\quad\;\;\; {\rm otherwise}
\end{array}\right.
\end{aligned}&&
\end{flalign*}
\begin{flalign*}
&\begin{aligned}
d=\left\{\begin{array}{rl}
0 & \quad\quad\;\;\; {\rm Ehrenfest} \\
1 & \quad\quad\;\;\; GD\;{\rm for\;electrons},\;{\rm 
bubble\;for\;bosons}
\end{array}\right.
\end{aligned}&&
\end{flalign*}
In Eq.~(\ref{GKBA-ODE5}) we have introduced
\begin{align}
\mathbfit{u}_{x}&\equiv (1-x)\,\mv+x\,\mw
\end{align}
and
\begin{align}
\mPsi^{e}(t)&\equiv 
  \mrho^{>}(t)\,\mathbfit{u}_{x}(t)\,\mrho^{<}(t)-\mrho^{<}(t)\,\mathbfit{u}_{x}(t) \,\mrho^{>}(t)
\end{align} 
to distinguish $GW$ and $T^{ph}$ ($x=0$) from  $GW+X$ and $T^{ph}+X$ 
($x=1$). Taking into account  Table~\ref{tab:1}
and the defintions of $h^{e}$, $\mh^{b}$ and $\mPsi^{b}$  in 
Eqs.~(\ref{eq:hele}), (\ref{hb}) and (\ref{Psi:b:def})
the whole NEGF toolbox is thus equivalent to a system of five coupled 
first-order ODE.

Equations~(\ref{GKBA-ODE}) fulfill all fundamental conservation laws and  
constitute the main result of this work.
As with any set of first-order differential equations the GKBA+ODE 
scheme must be supplied with an initial condition for the unknown 
quantities. We could start with an uncorrelated state 
described by a HF electronic density matrix and no bosons, i.e.,
\begin{subequations}
\begin{align}
\phi_{\mu}(0)&=   \langle 0| \hphi_{\mu} |0 \rangle,
\nonumber\\
\rho^{<}(0)&=\rho^{\rm HF},
\nonumber\\
\gamma^{<}_{\mu\nu}(0)&=\langle 0| \Delta \hphi_{\nu} \Delta 
\hphi_{\mu}|0 \rangle,
\nonumber\\
\mcG^{b}&=\mcG^{ee}=0,
\nonumber
\end{align}  
\end{subequations}
and then switch on the couplings $v(t)$ and $g(t)$ adiabatically. At 
the end of the adiabatic switching the values of 
$\phi$, $\rho^{<}$, $\bgg^{<}$, $\mcG^{ee}$ and $\mcG^{b}$ can be 
saved and used
as the initial (correlated) conditions for the simulations of 
interest.
However, we could also start from any initial nonequilibrium state.
In paper II we implement and solve Eqs.~(\ref{GKBA-ODE}) to  study the dynamics of polarons and 
phonon-dressed doublons in the Hubbard-Holstein model starting from 
two different nonequilibrium initial conditions, highlighting 
the effects of the interplay between $e$-$e$ and $e$-$b$ interactions.

We remark that the GKBA+ODE scheme is particularly advantageous to 
investigate  dynamical processes occurring at different time scales. 
Consider for instance the optical excitation in a semiconductor. Initially 
the dynamics is dictated by the electronic time-scale and hence the 
time-step to solve numerically Eqs.~(\ref{GKBA-ODE}) should be kept 
below $\sim$ 10 attoseconds. After a while the dynamics 
slows down and the time-scale is dictated by the electron-phonon 
scattering rate. We 
could then stop the simulation, save the value of $\phi$, $\rho^{<}$, 
$\bgg^{<}$, $\mcG^{ee}$ and $\mcG^{b}$, and start a second simulation 
using a larger time-step, e.g., $1\div 10$ femtoseconds, and the saved  values 
as initial conditions. Alternatively, we could implement an adaptive 
time-step. This second option is always preferable if we have no 
limits to the CPU time per job.


\section{Summary and outlook\label{sec:summary}}
In this work we developed a diagrammatic NEGF formalism to simulate
the coupled electron-boson dynamics in correlated materials. 
The formalism relies on the GKBA for 
electrons~\cite{lipavsky_generalized_1986} and on our recently 
proposed GKBA for bosons~\cite{karlsson_fast_2021}. With the {\mbox GKBAs}  one can 
collapse the KBE for the two-times Green's functions onto 
integro-differential equations 
for the one-time density matrices. 
In Refs.~\cite{schlunzen_achieving_2020, 
joost_g1-g2_2020} it was realized that the KBE+GKBA 
integro-differential equations of 
purely electronic systems can be reformulated in terms of a system of 
coupled first-order ODE 
for the $GW$ and $T$-matrix self-energies.
Shortly 
after such GKBA+ODE scheme was extended to include exchange 
effects and three-particle correlations~\cite{pavlyukh_photoinduced_2021}. Furthermore it was realized 
that a GKBA+ODE scheme can be constructed also to treat systems with only 
$e$-$b$ interactions~\cite{karlsson_fast_2021}. Our work shows
that $e$-$e$ and $e$-$b$ interactions can be treated on equal footing 
without altering the time-linear scaling.

We have presented a large class of methods, and for each of them the diagrammatic 
content of the self-energy has been explicitly worked out.
 The merits of the NEGF 
toolbox are (i) all fundamental conservation laws are satisfied 
independently of the  method; (2) the ODE nature of the EOM allows 
one to address phenomena occurring at different time scales through a 
save-and-restart procedure accompanied by an adaptation of the 
time-step; (3) as a by-product of the calculation we have access to 
the spatially non-local correlators $\mcG^{e}$ and $\mcG^{b}$, 
containing information on charge or magnetic orders~\cite{tuovinen_adiabatic_2019,tuovinen_comparing_2020}, polaronic 
or polaritonic states, etc., see Ref.~\cite{fehske_spatiotemporal_2011} and 
paper II. 

The formal development of the GKBA+ODE scheme is still at its 
infancy. The generalization of the GKBA to higher order Green's 
functions put forward in Ref.~\cite{pavlyukh_photoinduced_2021} may give access to even more accurate approximations 
while remaining within a time-linear scheme. Furthermore, the 
inclusion of an interaction with a fermionic or bosonic bath would 
make possible to simulate the dynamics of, e.g., photoionized 
systems~\cite{perfetto_ultrafast_2020,Adenine_2021} or 
molecular 
junctions~\cite{latini_charge_2014,tuovinen_electronic_2021}.
Numerical works based on the GKBA+ODE scheme have begun to appear in 
the literature only recently~\cite{pavlyukh_photoinduced_2021,perfetto_tdgw_2021,Borkowski_doublon_2021}. 
Parallel implementations in high performance computer facilities are 
expected to open the 
door to first-principles investigations of a wide 
range of nonequilibrium correlated phenomena.

\begin{acknowledgments}
  We acknowledge the financial support from MIUR PRIN (Grant No. 20173B72NB), from INFN
  through the TIME2QUEST project, and from Tor Vergata University through the Beyond
  Borders Project ULEXIEX.
\end{acknowledgments}


\begin{thebibliography}{68}%
\makeatletter
\providecommand \@ifxundefined [1]{%
 \@ifx{#1\undefined}
}%
\providecommand \@ifnum [1]{%
 \ifnum #1\expandafter \@firstoftwo
 \else \expandafter \@secondoftwo
 \fi
}%
\providecommand \@ifx [1]{%
 \ifx #1\expandafter \@firstoftwo
 \else \expandafter \@secondoftwo
 \fi
}%
\providecommand \natexlab [1]{#1}%
\providecommand \enquote  [1]{``#1''}%
\providecommand \bibnamefont  [1]{#1}%
\providecommand \bibfnamefont [1]{#1}%
\providecommand \citenamefont [1]{#1}%
\providecommand \href@noop [0]{\@secondoftwo}%
\providecommand \href [0]{\begingroup \@sanitize@url \@href}%
\providecommand \@href[1]{\@@startlink{#1}\@@href}%
\providecommand \@@href[1]{\endgroup#1\@@endlink}%
\providecommand \@sanitize@url [0]{\catcode `\\12\catcode `\$12\catcode
  `\&12\catcode `\#12\catcode `\^12\catcode `\_12\catcode `\%12\relax}%
\providecommand \@@startlink[1]{}%
\providecommand \@@endlink[0]{}%
\providecommand \url  [0]{\begingroup\@sanitize@url \@url }%
\providecommand \@url [1]{\endgroup\@href {#1}{\urlprefix }}%
\providecommand \urlprefix  [0]{URL }%
\providecommand \Eprint [0]{\href }%
\providecommand \doibase [0]{https://doi.org/}%
\providecommand \selectlanguage [0]{\@gobble}%
\providecommand \bibinfo  [0]{\@secondoftwo}%
\providecommand \bibfield  [0]{\@secondoftwo}%
\providecommand \translation [1]{[#1]}%
\providecommand \BibitemOpen [0]{}%
\providecommand \bibitemStop [0]{}%
\providecommand \bibitemNoStop [0]{.\EOS\space}%
\providecommand \EOS [0]{\spacefactor3000\relax}%
\providecommand \BibitemShut  [1]{\csname bibitem#1\endcsname}%
\let\auto@bib@innerbib\@empty
\bibitem [{\citenamefont {Ziman}(1960)}]{ziman_electrons_1960}%
  \BibitemOpen
  \bibfield  {author} {\bibinfo {author} {\bibfnamefont {J.~M.}\ \bibnamefont
  {Ziman}},\ }\href@noop {} {\emph {\bibinfo {title} {Electrons and Phonons:
  The Theory of Transport Phenomena in Solids}}}\ (\bibinfo  {publisher}
  {Oxford University Press},\ \bibinfo {address} {Oxford},\ \bibinfo {year}
  {1960})\BibitemShut {NoStop}%
\bibitem [{\citenamefont {Giustino}(2017)}]{giustino_electron-phonon_2017}%
  \BibitemOpen
  \bibfield  {author} {\bibinfo {author} {\bibfnamefont {F.}~\bibnamefont
  {Giustino}},\ }\bibfield  {title} {\bibinfo {title} {Electron-phonon
  interactions from first principles},\ }\href
  {https://doi.org/10.1103/RevModPhys.89.015003} {\bibfield  {journal}
  {\bibinfo  {journal} {Rev. Mod. Phys.}\ }\textbf {\bibinfo {volume} {89}},\
  \bibinfo {pages} {015003} (\bibinfo {year} {2017})}\BibitemShut {NoStop}%
\bibitem [{\citenamefont {Baroni}\ \emph {et~al.}(2001)\citenamefont {Baroni},
  \citenamefont {de~Gironcoli}, \citenamefont {Dal~Corso},\ and\ \citenamefont
  {Giannozzi}}]{baroni_phonons_2001}%
  \BibitemOpen
  \bibfield  {author} {\bibinfo {author} {\bibfnamefont {S.}~\bibnamefont
  {Baroni}}, \bibinfo {author} {\bibfnamefont {S.}~\bibnamefont
  {de~Gironcoli}}, \bibinfo {author} {\bibfnamefont {A.}~\bibnamefont
  {Dal~Corso}},\ and\ \bibinfo {author} {\bibfnamefont {P.}~\bibnamefont
  {Giannozzi}},\ }\bibfield  {title} {\bibinfo {title} {Phonons and related
  crystal properties from density-functional perturbation theory},\ }\href
  {https://doi.org/10.1103/RevModPhys.73.515} {\bibfield  {journal} {\bibinfo
  {journal} {Rev. Mod. Phys.}\ }\textbf {\bibinfo {volume} {73}},\ \bibinfo
  {pages} {515} (\bibinfo {year} {2001})}\BibitemShut {NoStop}%
\bibitem [{\citenamefont {Verdozzi}\ \emph {et~al.}(2006)\citenamefont
  {Verdozzi}, \citenamefont {Stefanucci},\ and\ \citenamefont
  {Almbladh}}]{verdozzi_classical_2006}%
  \BibitemOpen
  \bibfield  {author} {\bibinfo {author} {\bibfnamefont {C.}~\bibnamefont
  {Verdozzi}}, \bibinfo {author} {\bibfnamefont {G.}~\bibnamefont
  {Stefanucci}},\ and\ \bibinfo {author} {\bibfnamefont {C.-O.}\ \bibnamefont
  {Almbladh}},\ }\bibfield  {title} {\bibinfo {title} {Classical {Nuclear}
  {Motion} in {Quantum} {Transport}},\ }\href
  {https://doi.org/10.1103/PhysRevLett.97.046603} {\bibfield  {journal}
  {\bibinfo  {journal} {Phys. Rev. Lett.}\ }\textbf {\bibinfo {volume} {97}},\
  \bibinfo {pages} {046603} (\bibinfo {year} {2006})}\BibitemShut {NoStop}%
\bibitem [{\citenamefont {de~Melo}\ and\ \citenamefont
  {Marini}(2016)}]{de_melo_unified_2016}%
  \BibitemOpen
  \bibfield  {author} {\bibinfo {author} {\bibfnamefont {P.~M. M.~C.}\
  \bibnamefont {de~Melo}}\ and\ \bibinfo {author} {\bibfnamefont
  {A.}~\bibnamefont {Marini}},\ }\bibfield  {title} {\bibinfo {title} {Unified
  theory of quantized electrons, phonons, and photons out of equilibrium: {A}
  simplified ab initio approach based on the generalized {Baym}-{Kadanoff}
  ansatz},\ }\href {https://doi.org/10.1103/PhysRevB.93.155102} {\bibfield
  {journal} {\bibinfo  {journal} {Phys. Rev. B}\ }\textbf {\bibinfo {volume}
  {93}},\ \bibinfo {pages} {155102} (\bibinfo {year} {2016})}\BibitemShut
  {NoStop}%
\bibitem [{\citenamefont {Konstantinova}\ \emph {et~al.}(2018)\citenamefont
  {Konstantinova}, \citenamefont {Rameau}, \citenamefont {Reid}, \citenamefont
  {Abdurazakov}, \citenamefont {Wu}, \citenamefont {Li}, \citenamefont {Shen},
  \citenamefont {Gu}, \citenamefont {Huang}, \citenamefont {Rettig},
  \citenamefont {Avigo}, \citenamefont {Ligges}, \citenamefont {Freericks},
  \citenamefont {Kemper}, \citenamefont {D{\"u}rr}, \citenamefont
  {Bovensiepen}, \citenamefont {Johnson}, \citenamefont {Wang},\ and\
  \citenamefont {Zhu}}]{konstantinova_nonequilibrium_2018}%
  \BibitemOpen
  \bibfield  {author} {\bibinfo {author} {\bibfnamefont {T.}~\bibnamefont
  {Konstantinova}}, \bibinfo {author} {\bibfnamefont {J.~D.}\ \bibnamefont
  {Rameau}}, \bibinfo {author} {\bibfnamefont {A.~H.}\ \bibnamefont {Reid}},
  \bibinfo {author} {\bibfnamefont {O.}~\bibnamefont {Abdurazakov}}, \bibinfo
  {author} {\bibfnamefont {L.}~\bibnamefont {Wu}}, \bibinfo {author}
  {\bibfnamefont {R.}~\bibnamefont {Li}}, \bibinfo {author} {\bibfnamefont
  {X.}~\bibnamefont {Shen}}, \bibinfo {author} {\bibfnamefont {G.}~\bibnamefont
  {Gu}}, \bibinfo {author} {\bibfnamefont {Y.}~\bibnamefont {Huang}}, \bibinfo
  {author} {\bibfnamefont {L.}~\bibnamefont {Rettig}}, \bibinfo {author}
  {\bibfnamefont {I.}~\bibnamefont {Avigo}}, \bibinfo {author} {\bibfnamefont
  {M.}~\bibnamefont {Ligges}}, \bibinfo {author} {\bibfnamefont {J.~K.}\
  \bibnamefont {Freericks}}, \bibinfo {author} {\bibfnamefont {A.~F.}\
  \bibnamefont {Kemper}}, \bibinfo {author} {\bibfnamefont {H.~A.}\
  \bibnamefont {D{\"u}rr}}, \bibinfo {author} {\bibfnamefont {U.}~\bibnamefont
  {Bovensiepen}}, \bibinfo {author} {\bibfnamefont {P.~D.}\ \bibnamefont
  {Johnson}}, \bibinfo {author} {\bibfnamefont {X.}~\bibnamefont {Wang}},\ and\
  \bibinfo {author} {\bibfnamefont {Y.}~\bibnamefont {Zhu}},\ }\bibfield
  {title} {\bibinfo {title} {Nonequilibrium electron and lattice dynamics of
  strongly correlated {Bi$_2$Sr$_2$CaCu$_2$O$_{8+\delta}$} single crystals},\
  }\href {https://www.science.org/doi/abs/10.1126/sciadv.aap7427} {\bibfield
  {journal} {\bibinfo  {journal} {Science Advances}\ }\textbf {\bibinfo
  {volume} {4}},\ \bibinfo {pages} {eaap7427} (\bibinfo {year}
  {2018})}\BibitemShut {NoStop}%
\bibitem [{\citenamefont {Rizzi}\ \emph {et~al.}(2016)\citenamefont {Rizzi},
  \citenamefont {Todorov}, \citenamefont {Kohanoff},\ and\ \citenamefont
  {Correa}}]{rizzi_electron-phonon_2016}%
  \BibitemOpen
  \bibfield  {author} {\bibinfo {author} {\bibfnamefont {V.}~\bibnamefont
  {Rizzi}}, \bibinfo {author} {\bibfnamefont {T.~N.}\ \bibnamefont {Todorov}},
  \bibinfo {author} {\bibfnamefont {J.~J.}\ \bibnamefont {Kohanoff}},\ and\
  \bibinfo {author} {\bibfnamefont {A.~A.}\ \bibnamefont {Correa}},\ }\bibfield
   {title} {\bibinfo {title} {Electron-phonon thermalization in a scalable
  method for real-time quantum dynamics},\ }\href
  {https://link.aps.org/doi/10.1103/PhysRevB.93.024306} {\bibfield  {journal}
  {\bibinfo  {journal} {Phys. Rev. B}\ }\textbf {\bibinfo {volume} {93}},\
  \bibinfo {pages} {024306} (\bibinfo {year} {2016})}\BibitemShut {NoStop}%
\bibitem [{\citenamefont {van Hest}\ \emph {et~al.}(2018)\citenamefont {van
  Hest}, \citenamefont {Blab}, \citenamefont {Gerritsen}, \citenamefont
  {de~Mello~Donega},\ and\ \citenamefont {Meijerink}}]{van_hest_the_role_2018}%
  \BibitemOpen
  \bibfield  {author} {\bibinfo {author} {\bibfnamefont {J.~J. H.~A.}\
  \bibnamefont {van Hest}}, \bibinfo {author} {\bibfnamefont {G.~A.}\
  \bibnamefont {Blab}}, \bibinfo {author} {\bibfnamefont {H.~C.}\ \bibnamefont
  {Gerritsen}}, \bibinfo {author} {\bibfnamefont {C.}~\bibnamefont
  {de~Mello~Donega}},\ and\ \bibinfo {author} {\bibfnamefont {A.}~\bibnamefont
  {Meijerink}},\ }\bibfield  {title} {\bibinfo {title} {The role of a phonon
  bottleneck in relaxation processes for ln-doped nayf$_4$ nanocrystals},\
  }\href {https://doi.org/10.1021/acs.jpcc.7b11171} {\bibfield  {journal}
  {\bibinfo  {journal} {The Journal of Physical Chemistry C}\ }\textbf
  {\bibinfo {volume} {122}},\ \bibinfo {pages} {3985} (\bibinfo {year}
  {2018})}\BibitemShut {NoStop}%
\bibitem [{\citenamefont {Ruggenthaler}\ \emph {et~al.}(2018)\citenamefont
  {Ruggenthaler}, \citenamefont {Tancogne-Dejean}, \citenamefont {Flick},
  \citenamefont {Appel},\ and\ \citenamefont
  {Rubio}}]{ruggenthaler_quantum-electrodynamical_2018}%
  \BibitemOpen
  \bibfield  {author} {\bibinfo {author} {\bibfnamefont {M.}~\bibnamefont
  {Ruggenthaler}}, \bibinfo {author} {\bibfnamefont {N.}~\bibnamefont
  {Tancogne-Dejean}}, \bibinfo {author} {\bibfnamefont {J.}~\bibnamefont
  {Flick}}, \bibinfo {author} {\bibfnamefont {H.}~\bibnamefont {Appel}},\ and\
  \bibinfo {author} {\bibfnamefont {A.}~\bibnamefont {Rubio}},\ }\bibfield
  {title} {\bibinfo {title} {From a quantum-electrodynamical light-matter
  description to novel spectroscopies},\ }\href
  {https://doi.org/10.1038/s41570-018-0118} {\bibfield  {journal} {\bibinfo
  {journal} {Nature Reviews Chemistry}\ }\textbf {\bibinfo {volume} {2}},\
  \bibinfo {pages} {0118} (\bibinfo {year} {2018})}\BibitemShut {NoStop}%
\bibitem [{\citenamefont {Wang}\ \emph {et~al.}(2019)\citenamefont {Wang},
  \citenamefont {Chen}, \citenamefont {Liang}, \citenamefont {Li},
  \citenamefont {Lai}, \citenamefont {Ding},\ and\ \citenamefont
  {Wu}}]{wang_observation_2019}%
  \BibitemOpen
  \bibfield  {author} {\bibinfo {author} {\bibfnamefont {L.}~\bibnamefont
  {Wang}}, \bibinfo {author} {\bibfnamefont {Z.}~\bibnamefont {Chen}}, \bibinfo
  {author} {\bibfnamefont {G.}~\bibnamefont {Liang}}, \bibinfo {author}
  {\bibfnamefont {Y.}~\bibnamefont {Li}}, \bibinfo {author} {\bibfnamefont
  {R.}~\bibnamefont {Lai}}, \bibinfo {author} {\bibfnamefont {T.}~\bibnamefont
  {Ding}},\ and\ \bibinfo {author} {\bibfnamefont {K.}~\bibnamefont {Wu}},\
  }\bibfield  {title} {\bibinfo {title} {Observation of a phonon bottleneck in
  copper-doped colloidal quantum dots},\ }\href
  {https://doi.org/10.1038/s41467-019-12558-y} {\bibfield  {journal} {\bibinfo
  {journal} {Nature Communications}\ }\textbf {\bibinfo {volume} {10}},\
  \bibinfo {pages} {4532} (\bibinfo {year} {2019})}\BibitemShut {NoStop}%
\bibitem [{\citenamefont {Karlsson}\ \emph {et~al.}(2021)\citenamefont
  {Karlsson}, \citenamefont {van Leeuwen}, \citenamefont {Pavlyukh},
  \citenamefont {Perfetto},\ and\ \citenamefont
  {Stefanucci}}]{karlsson_fast_2021}%
  \BibitemOpen
  \bibfield  {author} {\bibinfo {author} {\bibfnamefont {D.}~\bibnamefont
  {Karlsson}}, \bibinfo {author} {\bibfnamefont {R.}~\bibnamefont {van
  Leeuwen}}, \bibinfo {author} {\bibfnamefont {Y.}~\bibnamefont {Pavlyukh}},
  \bibinfo {author} {\bibfnamefont {E.}~\bibnamefont {Perfetto}},\ and\
  \bibinfo {author} {\bibfnamefont {G.}~\bibnamefont {Stefanucci}},\ }\bibfield
   {title} {\bibinfo {title} {Fast {Green}'s {Function} {Method} for
  {Ultrafast} {Electron}-{Boson} {Dynamics}},\ }\href
  {https://doi.org/10.1103/PhysRevLett.127.036402} {\bibfield  {journal}
  {\bibinfo  {journal} {Phys. Rev. Lett.}\ }\textbf {\bibinfo {volume} {127}},\
  \bibinfo {pages} {036402} (\bibinfo {year} {2021})}\BibitemShut {NoStop}%
\bibitem [{\citenamefont {Mankowsky}\ \emph {et~al.}(2014)\citenamefont
  {Mankowsky}, \citenamefont {Subedi}, \citenamefont {F\"{o}rst}, \citenamefont
  {Mariager}, \citenamefont {Chollet}, \citenamefont {Lemke}, \citenamefont
  {Robinson}, \citenamefont {Glownia}, \citenamefont {Minitti}, \citenamefont
  {Frano}, \citenamefont {Fechner}, \citenamefont {Spaldin}, \citenamefont
  {Loew}, \citenamefont {Keimer}, \citenamefont {Georges},\ and\ \citenamefont
  {Cavalleri}}]{mankowsky_nonlinear_2014}%
  \BibitemOpen
  \bibfield  {author} {\bibinfo {author} {\bibfnamefont {R.}~\bibnamefont
  {Mankowsky}}, \bibinfo {author} {\bibfnamefont {A.}~\bibnamefont {Subedi}},
  \bibinfo {author} {\bibfnamefont {M.}~\bibnamefont {F\"{o}rst}}, \bibinfo
  {author} {\bibfnamefont {S.~O.}\ \bibnamefont {Mariager}}, \bibinfo {author}
  {\bibfnamefont {M.}~\bibnamefont {Chollet}}, \bibinfo {author} {\bibfnamefont
  {H.~T.}\ \bibnamefont {Lemke}}, \bibinfo {author} {\bibfnamefont {J.~S.}\
  \bibnamefont {Robinson}}, \bibinfo {author} {\bibfnamefont {J.~M.}\
  \bibnamefont {Glownia}}, \bibinfo {author} {\bibfnamefont {M.~P.}\
  \bibnamefont {Minitti}}, \bibinfo {author} {\bibfnamefont {A.}~\bibnamefont
  {Frano}}, \bibinfo {author} {\bibfnamefont {M.}~\bibnamefont {Fechner}},
  \bibinfo {author} {\bibfnamefont {N.~A.}\ \bibnamefont {Spaldin}}, \bibinfo
  {author} {\bibfnamefont {T.}~\bibnamefont {Loew}}, \bibinfo {author}
  {\bibfnamefont {B.}~\bibnamefont {Keimer}}, \bibinfo {author} {\bibfnamefont
  {A.}~\bibnamefont {Georges}},\ and\ \bibinfo {author} {\bibfnamefont
  {A.}~\bibnamefont {Cavalleri}},\ }\bibfield  {title} {\bibinfo {title}
  {Nonlinear lattice dynamics as a basis for enhanced superconductivity in
  {YBa2Cu3O6}.5},\ }\href {https://doi.org/10.1038/nature13875} {\bibfield
  {journal} {\bibinfo  {journal} {Nature}\ }\textbf {\bibinfo {volume} {516}},\
  \bibinfo {pages} {71} (\bibinfo {year} {2014})}\BibitemShut {NoStop}%
\bibitem [{\citenamefont {Mitrano}\ \emph {et~al.}(2016)\citenamefont
  {Mitrano}, \citenamefont {Cantaluppi}, \citenamefont {Nicoletti},
  \citenamefont {Kaiser}, \citenamefont {Perucchi}, \citenamefont {Lupi},
  \citenamefont {Di~Pietro}, \citenamefont {Pontiroli}, \citenamefont
  {Ricc\`{o}}, \citenamefont {Clark}, \citenamefont {Jaksch},\ and\
  \citenamefont {Cavalleri}}]{mitrano_possible_2016}%
  \BibitemOpen
  \bibfield  {author} {\bibinfo {author} {\bibfnamefont {M.}~\bibnamefont
  {Mitrano}}, \bibinfo {author} {\bibfnamefont {A.}~\bibnamefont {Cantaluppi}},
  \bibinfo {author} {\bibfnamefont {D.}~\bibnamefont {Nicoletti}}, \bibinfo
  {author} {\bibfnamefont {S.}~\bibnamefont {Kaiser}}, \bibinfo {author}
  {\bibfnamefont {A.}~\bibnamefont {Perucchi}}, \bibinfo {author}
  {\bibfnamefont {S.}~\bibnamefont {Lupi}}, \bibinfo {author} {\bibfnamefont
  {P.}~\bibnamefont {Di~Pietro}}, \bibinfo {author} {\bibfnamefont
  {D.}~\bibnamefont {Pontiroli}}, \bibinfo {author} {\bibfnamefont
  {M.}~\bibnamefont {Ricc\`{o}}}, \bibinfo {author} {\bibfnamefont {S.~R.}\
  \bibnamefont {Clark}}, \bibinfo {author} {\bibfnamefont {D.}~\bibnamefont
  {Jaksch}},\ and\ \bibinfo {author} {\bibfnamefont {A.}~\bibnamefont
  {Cavalleri}},\ }\bibfield  {title} {\bibinfo {title} {Possible light-induced
  superconductivity in {K3C60} at high temperature},\ }\href
  {https://doi.org/10.1038/nature16522} {\bibfield  {journal} {\bibinfo
  {journal} {Nature}\ }\textbf {\bibinfo {volume} {530}},\ \bibinfo {pages}
  {461} (\bibinfo {year} {2016})}\BibitemShut {NoStop}%
\bibitem [{\citenamefont {Sentef}\ \emph {et~al.}(2016)\citenamefont {Sentef},
  \citenamefont {Kemper}, \citenamefont {Georges},\ and\ \citenamefont
  {Kollath}}]{sentef_theory_2016}%
  \BibitemOpen
  \bibfield  {author} {\bibinfo {author} {\bibfnamefont {M.~A.}\ \bibnamefont
  {Sentef}}, \bibinfo {author} {\bibfnamefont {A.~F.}\ \bibnamefont {Kemper}},
  \bibinfo {author} {\bibfnamefont {A.}~\bibnamefont {Georges}},\ and\ \bibinfo
  {author} {\bibfnamefont {C.}~\bibnamefont {Kollath}},\ }\bibfield  {title}
  {\bibinfo {title} {Theory of light-enhanced phonon-mediated
  superconductivity},\ }\href {https://doi.org/10.1103/PhysRevB.93.144506}
  {\bibfield  {journal} {\bibinfo  {journal} {Phys. Rev. B}\ }\textbf {\bibinfo
  {volume} {93}},\ \bibinfo {pages} {144506} (\bibinfo {year}
  {2016})}\BibitemShut {NoStop}%
\bibitem [{\citenamefont {Babadi}\ \emph {et~al.}(2017)\citenamefont {Babadi},
  \citenamefont {Knap}, \citenamefont {Martin}, \citenamefont {Refael},\ and\
  \citenamefont {Demler}}]{babadi_theory_2017}%
  \BibitemOpen
  \bibfield  {author} {\bibinfo {author} {\bibfnamefont {M.}~\bibnamefont
  {Babadi}}, \bibinfo {author} {\bibfnamefont {M.}~\bibnamefont {Knap}},
  \bibinfo {author} {\bibfnamefont {I.}~\bibnamefont {Martin}}, \bibinfo
  {author} {\bibfnamefont {G.}~\bibnamefont {Refael}},\ and\ \bibinfo {author}
  {\bibfnamefont {E.}~\bibnamefont {Demler}},\ }\bibfield  {title} {\bibinfo
  {title} {Theory of parametrically amplified electron-phonon
  superconductivity},\ }\href {https://doi.org/10.1103/PhysRevB.96.014512}
  {\bibfield  {journal} {\bibinfo  {journal} {Phys. Rev. B}\ }\textbf {\bibinfo
  {volume} {96}},\ \bibinfo {pages} {014512} (\bibinfo {year}
  {2017})}\BibitemShut {NoStop}%
\bibitem [{\citenamefont {Gudmundsson}\ \emph {et~al.}(2012)\citenamefont
  {Gudmundsson}, \citenamefont {Jonasson}, \citenamefont {Tang}, \citenamefont
  {Goan},\ and\ \citenamefont {Manolescu}}]{gudmundsson_time-dependent_2012}%
  \BibitemOpen
  \bibfield  {author} {\bibinfo {author} {\bibfnamefont {V.}~\bibnamefont
  {Gudmundsson}}, \bibinfo {author} {\bibfnamefont {O.}~\bibnamefont
  {Jonasson}}, \bibinfo {author} {\bibfnamefont {C.-S.}\ \bibnamefont {Tang}},
  \bibinfo {author} {\bibfnamefont {H.-S.}\ \bibnamefont {Goan}},\ and\
  \bibinfo {author} {\bibfnamefont {A.}~\bibnamefont {Manolescu}},\ }\bibfield
  {title} {\bibinfo {title} {Time-dependent transport of electrons through a
  photon cavity},\ }\href {https://doi.org/10.1103/PhysRevB.85.075306}
  {\bibfield  {journal} {\bibinfo  {journal} {Phys. Rev. B}\ }\textbf {\bibinfo
  {volume} {85}},\ \bibinfo {pages} {075306} (\bibinfo {year}
  {2012})}\BibitemShut {NoStop}%
\bibitem [{\citenamefont {Abdullah}\ \emph {et~al.}(2018)\citenamefont
  {Abdullah}, \citenamefont {Tang}, \citenamefont {Manolescu},\ and\
  \citenamefont {Gudmundsson}}]{abdullah_effects_2018}%
  \BibitemOpen
  \bibfield  {author} {\bibinfo {author} {\bibfnamefont {N.~R.}\ \bibnamefont
  {Abdullah}}, \bibinfo {author} {\bibfnamefont {C.-S.}\ \bibnamefont {Tang}},
  \bibinfo {author} {\bibfnamefont {A.}~\bibnamefont {Manolescu}},\ and\
  \bibinfo {author} {\bibfnamefont {V.}~\bibnamefont {Gudmundsson}},\
  }\bibfield  {title} {\bibinfo {title} {Effects of photon field on heat
  transport through a quantum wire attached to leads},\ }\href
  {https://www.sciencedirect.com/science/article/pii/S0375960117311209}
  {\bibfield  {journal} {\bibinfo  {journal} {Physics Letters A}\ }\textbf
  {\bibinfo {volume} {382}},\ \bibinfo {pages} {199} (\bibinfo {year}
  {2018})}\BibitemShut {NoStop}%
\bibitem [{\citenamefont {Galperin}(2017)}]{galperin_photonics_2017}%
  \BibitemOpen
  \bibfield  {author} {\bibinfo {author} {\bibfnamefont {M.}~\bibnamefont
  {Galperin}},\ }\bibfield  {title} {\bibinfo {title} {Photonics and
  spectroscopy in nanojunctions: a theoretical insight},\ }\href
  {https://doi.org/10.1039/C7CS00067G} {\bibfield  {journal} {\bibinfo
  {journal} {Chem. Soc. Rev.}\ }\textbf {\bibinfo {volume} {46}},\ \bibinfo
  {pages} {4000} (\bibinfo {year} {2017})}\BibitemShut {NoStop}%
\bibitem [{\citenamefont {Chen}\ \emph {et~al.}(2020)\citenamefont {Chen},
  \citenamefont {Sangalli},\ and\ \citenamefont
  {Bernardi}}]{chen_exciton-phonon_2020}%
  \BibitemOpen
  \bibfield  {author} {\bibinfo {author} {\bibfnamefont {H.-Y.}\ \bibnamefont
  {Chen}}, \bibinfo {author} {\bibfnamefont {D.}~\bibnamefont {Sangalli}},\
  and\ \bibinfo {author} {\bibfnamefont {M.}~\bibnamefont {Bernardi}},\
  }\bibfield  {title} {\bibinfo {title} {Exciton-phonon interaction and
  relaxation times from first principles},\ }\href
  {https://doi.org/10.1103/PhysRevLett.125.107401} {\bibfield  {journal}
  {\bibinfo  {journal} {Phys. Rev. Lett.}\ }\textbf {\bibinfo {volume} {125}},\
  \bibinfo {pages} {107401} (\bibinfo {year} {2020})}\BibitemShut {NoStop}%
\bibitem [{\citenamefont {Stefanucci}\ and\ \citenamefont
  {Perfetto}(2021)}]{stefanucci_carriers_2021}%
  \BibitemOpen
  \bibfield  {author} {\bibinfo {author} {\bibfnamefont {G.}~\bibnamefont
  {Stefanucci}}\ and\ \bibinfo {author} {\bibfnamefont {E.}~\bibnamefont
  {Perfetto}},\ }\bibfield  {title} {\bibinfo {title} {From carriers and
  virtual excitons to exciton populations: {Insights} into time-resolved
  {ARPES} spectra from an exactly solvable model},\ }\href
  {https://doi.org/10.1103/PhysRevB.103.245103} {\bibfield  {journal} {\bibinfo
   {journal} {Phys. Rev. B}\ }\textbf {\bibinfo {volume} {103}},\ \bibinfo
  {pages} {245103} (\bibinfo {year} {2021})}\BibitemShut {NoStop}%
\bibitem [{\citenamefont {Helmrich}\ \emph {et~al.}(2021)\citenamefont
  {Helmrich}, \citenamefont {Sampson}, \citenamefont {Huang}, \citenamefont
  {Selig}, \citenamefont {Hao}, \citenamefont {Tran}, \citenamefont {Achstein},
  \citenamefont {Young}, \citenamefont {Knorr}, \citenamefont {Malic},
  \citenamefont {Woggon}, \citenamefont {Owschimikow},\ and\ \citenamefont
  {Li}}]{helmrich_phonon-assisted_2021}%
  \BibitemOpen
  \bibfield  {author} {\bibinfo {author} {\bibfnamefont {S.}~\bibnamefont
  {Helmrich}}, \bibinfo {author} {\bibfnamefont {K.}~\bibnamefont {Sampson}},
  \bibinfo {author} {\bibfnamefont {D.}~\bibnamefont {Huang}}, \bibinfo
  {author} {\bibfnamefont {M.}~\bibnamefont {Selig}}, \bibinfo {author}
  {\bibfnamefont {K.}~\bibnamefont {Hao}}, \bibinfo {author} {\bibfnamefont
  {K.}~\bibnamefont {Tran}}, \bibinfo {author} {\bibfnamefont {A.}~\bibnamefont
  {Achstein}}, \bibinfo {author} {\bibfnamefont {C.}~\bibnamefont {Young}},
  \bibinfo {author} {\bibfnamefont {A.}~\bibnamefont {Knorr}}, \bibinfo
  {author} {\bibfnamefont {E.}~\bibnamefont {Malic}}, \bibinfo {author}
  {\bibfnamefont {U.}~\bibnamefont {Woggon}}, \bibinfo {author} {\bibfnamefont
  {N.}~\bibnamefont {Owschimikow}},\ and\ \bibinfo {author} {\bibfnamefont
  {X.}~\bibnamefont {Li}},\ }\bibfield  {title} {\bibinfo {title}
  {Phonon-assisted intervalley scattering determines ultrafast exciton dynamics
  in ${\mathrm{mose}}_{2}$ bilayers},\ }\href
  {https://doi.org/10.1103/PhysRevLett.127.157403} {\bibfield  {journal}
  {\bibinfo  {journal} {Phys. Rev. Lett.}\ }\textbf {\bibinfo {volume} {127}},\
  \bibinfo {pages} {157403} (\bibinfo {year} {2021})}\BibitemShut {NoStop}%
\bibitem [{\citenamefont {Walther}\ \emph {et~al.}(2006)\citenamefont
  {Walther}, \citenamefont {Varcoe}, \citenamefont {Englert},\ and\
  \citenamefont {Becker}}]{walther_cavity_2006}%
  \BibitemOpen
  \bibfield  {author} {\bibinfo {author} {\bibfnamefont {H.}~\bibnamefont
  {Walther}}, \bibinfo {author} {\bibfnamefont {B.~T.~H.}\ \bibnamefont
  {Varcoe}}, \bibinfo {author} {\bibfnamefont {B.-G.}\ \bibnamefont
  {Englert}},\ and\ \bibinfo {author} {\bibfnamefont {T.}~\bibnamefont
  {Becker}},\ }\bibfield  {title} {\bibinfo {title} {Cavity quantum
  electrodynamics},\ }\href {https://doi.org/10.1088/0034-4885/69/5/r02}
  {\bibfield  {journal} {\bibinfo  {journal} {Report Progrosses in Physics}\
  }\textbf {\bibinfo {volume} {69}},\ \bibinfo {pages} {1325} (\bibinfo {year}
  {2006})}\BibitemShut {NoStop}%
\bibitem [{\citenamefont {Hutchison}\ \emph {et~al.}(2012)\citenamefont
  {Hutchison}, \citenamefont {Schwartz}, \citenamefont {Genet}, \citenamefont
  {Devaux},\ and\ \citenamefont {Ebbesen}}]{hutchison_modifying_2012}%
  \BibitemOpen
  \bibfield  {author} {\bibinfo {author} {\bibfnamefont {J.~A.}\ \bibnamefont
  {Hutchison}}, \bibinfo {author} {\bibfnamefont {T.}~\bibnamefont {Schwartz}},
  \bibinfo {author} {\bibfnamefont {C.}~\bibnamefont {Genet}}, \bibinfo
  {author} {\bibfnamefont {E.}~\bibnamefont {Devaux}},\ and\ \bibinfo {author}
  {\bibfnamefont {T.~W.}\ \bibnamefont {Ebbesen}},\ }\bibfield  {title}
  {\bibinfo {title} {Modifying chemical landscapes by coupling to vacuum
  fields},\ }\href
  {https://onlinelibrary.wiley.com/doi/abs/10.1002/anie.201107033} {\bibfield
  {journal} {\bibinfo  {journal} {Angewandte Chemie International Edition}\
  }\textbf {\bibinfo {volume} {51}},\ \bibinfo {pages} {1592} (\bibinfo {year}
  {2012})}\BibitemShut {NoStop}%
\bibitem [{\citenamefont {Danielewicz}(1984)}]{danielewicz_quantum_1984}%
  \BibitemOpen
  \bibfield  {author} {\bibinfo {author} {\bibfnamefont {P.}~\bibnamefont
  {Danielewicz}},\ }\bibfield  {title} {\bibinfo {title} {Quantum theory of
  nonequilibrium processes, {I}},\ }\href
  {https://doi.org/10.1016/0003-4916(84)90092-7} {\bibfield  {journal}
  {\bibinfo  {journal} {Ann. Phys.}\ }\textbf {\bibinfo {volume} {152}},\
  \bibinfo {pages} {239} (\bibinfo {year} {1984})}\BibitemShut {NoStop}%
\bibitem [{\citenamefont {van Leeuwen}\ \emph {et~al.}(2006)\citenamefont {van
  Leeuwen}, \citenamefont {Dahlen}, \citenamefont {Stefanucci}, \citenamefont
  {Almbladh},\ and\ \citenamefont {von Barth}}]{van_leeuwen_introduction_2006}%
  \BibitemOpen
  \bibfield  {author} {\bibinfo {author} {\bibfnamefont {R.}~\bibnamefont {van
  Leeuwen}}, \bibinfo {author} {\bibfnamefont {N.}~\bibnamefont {Dahlen}},
  \bibinfo {author} {\bibfnamefont {G.}~\bibnamefont {Stefanucci}}, \bibinfo
  {author} {\bibfnamefont {C.-O.}\ \bibnamefont {Almbladh}},\ and\ \bibinfo
  {author} {\bibfnamefont {U.}~\bibnamefont {von Barth}},\ }\bibfield  {title}
  {\bibinfo {title} {Introduction to the {Keldysh} {Formalism}},\ }in\ \href
  {http://www.springerlink.com/content/d40076h8251508l7/abstract/} {\emph
  {\bibinfo {booktitle} {Time-{Dependent} {Density} {Functional} {Theory}}}},\
  \bibinfo {series} {Lecture {Notes} in {Physics}}, Vol.\ \bibinfo {volume}
  {706},\ \bibinfo {editor} {edited by\ \bibinfo {editor} {\bibfnamefont
  {M.}~\bibnamefont {Marques}}, \bibinfo {editor} {\bibfnamefont
  {C.}~\bibnamefont {Ullrich}}, \bibinfo {editor} {\bibfnamefont
  {F.}~\bibnamefont {Nogueira}}, \bibinfo {editor} {\bibfnamefont
  {A.}~\bibnamefont {Rubio}}, \bibinfo {editor} {\bibfnamefont
  {K.}~\bibnamefont {Burke}},\ and\ \bibinfo {editor} {\bibfnamefont
  {E.}~\bibnamefont {Gross}}}\ (\bibinfo  {publisher} {Springer Berlin /
  Heidelberg},\ \bibinfo {year} {2006})\ pp.\ \bibinfo {pages}
  {33--59}\BibitemShut {NoStop}%
\bibitem [{\citenamefont {Stefanucci}\ and\ \citenamefont {van
  Leeuwen}(2013)}]{stefanucci_nonequilibrium_2013}%
  \BibitemOpen
  \bibfield  {author} {\bibinfo {author} {\bibfnamefont {G.}~\bibnamefont
  {Stefanucci}}\ and\ \bibinfo {author} {\bibfnamefont {R.}~\bibnamefont {van
  Leeuwen}},\ }\href {http://dx.doi.org/10.1017/CBO9781139023979} {\emph
  {\bibinfo {title} {Nonequilibrium {Many}-{Body} {Theory} of {Quantum}
  {Systems}: {A} {Modern} {Introduction}}}}\ (\bibinfo  {publisher} {Cambridge
  University Press},\ \bibinfo {address} {Cambridge},\ \bibinfo {year}
  {2013})\BibitemShut {NoStop}%
\bibitem [{\citenamefont {Karlsson}\ and\ \citenamefont {van
  Leeuwen}(2020)}]{karlsson_non-equilibrium_2018}%
  \BibitemOpen
  \bibfield  {author} {\bibinfo {author} {\bibfnamefont {D.}~\bibnamefont
  {Karlsson}}\ and\ \bibinfo {author} {\bibfnamefont {R.}~\bibnamefont {van
  Leeuwen}},\ }\bibinfo {title} {Non-equilibrium green's functions for coupled
  fermion-boson systems},\ in\ \href
  {https://doi.org/10.1007/978-3-319-44677-6_8} {\emph {\bibinfo {booktitle}
  {Handbook of Materials Modeling: Methods: Theory and Modeling}}},\ \bibinfo
  {editor} {edited by\ \bibinfo {editor} {\bibfnamefont {W.}~\bibnamefont
  {Andreoni}}\ and\ \bibinfo {editor} {\bibfnamefont {S.}~\bibnamefont {Yip}}}\
  (\bibinfo  {publisher} {Springer International Publishing},\ \bibinfo
  {address} {Cham},\ \bibinfo {year} {2020})\ pp.\ \bibinfo {pages}
  {367--395}\BibitemShut {NoStop}%
\bibitem [{\citenamefont {Kadanoff}\ and\ \citenamefont
  {Baym}(1962)}]{kadanoff_quantum_1962}%
  \BibitemOpen
  \bibfield  {author} {\bibinfo {author} {\bibfnamefont {L.}~\bibnamefont
  {Kadanoff}}\ and\ \bibinfo {author} {\bibfnamefont {G.}~\bibnamefont
  {Baym}},\ }\href@noop {} {\emph {\bibinfo {title} {Quantum statistical
  mechanics {Green}'s function methods in equilibrium and nonequilibrium
  problems}}}\ (\bibinfo  {publisher} {W.A. Benjamin},\ \bibinfo {address} {New
  York},\ \bibinfo {year} {1962})\BibitemShut {NoStop}%
\bibitem [{\citenamefont {Kwong}\ and\ \citenamefont
  {Bonitz}(2000)}]{kwong_real-time_2000}%
  \BibitemOpen
  \bibfield  {author} {\bibinfo {author} {\bibfnamefont {N.-H.}\ \bibnamefont
  {Kwong}}\ and\ \bibinfo {author} {\bibfnamefont {M.}~\bibnamefont {Bonitz}},\
  }\bibfield  {title} {\bibinfo {title} {Real-{Time} {Kadanoff}-{Baym}
  {Approach} to {Plasma} {Oscillations} in a {Correlated} {Electron} {Gas}},\
  }\href {https://doi.org/10.1103/PhysRevLett.84.1768} {\bibfield  {journal}
  {\bibinfo  {journal} {Phys. Rev. Lett.}\ }\textbf {\bibinfo {volume} {84}},\
  \bibinfo {pages} {1768} (\bibinfo {year} {2000})}\BibitemShut {NoStop}%
\bibitem [{\citenamefont {Dahlen}\ and\ \citenamefont {van
  Leeuwen}(2007)}]{dahlen_solving_2007}%
  \BibitemOpen
  \bibfield  {author} {\bibinfo {author} {\bibfnamefont {N.~E.}\ \bibnamefont
  {Dahlen}}\ and\ \bibinfo {author} {\bibfnamefont {R.}~\bibnamefont {van
  Leeuwen}},\ }\bibfield  {title} {\bibinfo {title} {Solving the
  {Kadanoff}-{Baym} {Equations} for {Inhomogeneous} {Systems}: {Application} to
  {Atoms} and {Molecules}},\ }\href
  {http://link.aps.org/doi/10.1103/PhysRevLett.98.153004} {\bibfield  {journal}
  {\bibinfo  {journal} {Phys. Rev. Lett.}\ }\textbf {\bibinfo {volume} {98}},\
  \bibinfo {pages} {153004} (\bibinfo {year} {2007})}\BibitemShut {NoStop}%
\bibitem [{\citenamefont {My\"{o}h\"{a}nen}\ \emph {et~al.}(2008)\citenamefont
  {My\"{o}h\"{a}nen}, \citenamefont {Stan}, \citenamefont {Stefanucci},\ and\
  \citenamefont {van Leeuwen}}]{myohanen_many-body_2008}%
  \BibitemOpen
  \bibfield  {author} {\bibinfo {author} {\bibfnamefont {P.}~\bibnamefont
  {My\"{o}h\"{a}nen}}, \bibinfo {author} {\bibfnamefont {A.}~\bibnamefont
  {Stan}}, \bibinfo {author} {\bibfnamefont {G.}~\bibnamefont {Stefanucci}},\
  and\ \bibinfo {author} {\bibfnamefont {R.}~\bibnamefont {van Leeuwen}},\
  }\bibfield  {title} {\bibinfo {title} {A many-body approach to quantum
  transport dynamics: {Initial} correlations and memory effects},\ }\href
  {https://doi.org/10.1209/0295-5075/84/67001} {\bibfield  {journal} {\bibinfo
  {journal} {Eurphys. Lett.}\ }\textbf {\bibinfo {volume} {84}},\ \bibinfo
  {pages} {67001} (\bibinfo {year} {2008})}\BibitemShut {NoStop}%
\bibitem [{\citenamefont {Galperin}\ and\ \citenamefont
  {Tretiak}(2008)}]{galperin_linear_2008}%
  \BibitemOpen
  \bibfield  {author} {\bibinfo {author} {\bibfnamefont {M.}~\bibnamefont
  {Galperin}}\ and\ \bibinfo {author} {\bibfnamefont {S.}~\bibnamefont
  {Tretiak}},\ }\bibfield  {title} {\bibinfo {title} {Linear optical response
  of current-carrying molecular junction: {A} nonequilibrium {Green}'s
  function-time-dependent density functional theory approach},\ }\href
  {https://doi.org/10.1063/1.2876011} {\bibfield  {journal} {\bibinfo
  {journal} {J. Chem. Phys.}\ }\textbf {\bibinfo {volume} {128}},\ \bibinfo
  {pages} {124705} (\bibinfo {year} {2008})}\BibitemShut {NoStop}%
\bibitem [{\citenamefont {My\"{o}h\"{a}nen}\ \emph {et~al.}(2009)\citenamefont
  {My\"{o}h\"{a}nen}, \citenamefont {Stan}, \citenamefont {Stefanucci},\ and\
  \citenamefont {van Leeuwen}}]{myohanen_kadanoff-baym_2009}%
  \BibitemOpen
  \bibfield  {author} {\bibinfo {author} {\bibfnamefont {P.}~\bibnamefont
  {My\"{o}h\"{a}nen}}, \bibinfo {author} {\bibfnamefont {A.}~\bibnamefont
  {Stan}}, \bibinfo {author} {\bibfnamefont {G.}~\bibnamefont {Stefanucci}},\
  and\ \bibinfo {author} {\bibfnamefont {R.}~\bibnamefont {van Leeuwen}},\
  }\bibfield  {title} {\bibinfo {title} {Kadanoff-{Baym} approach to quantum
  transport through interacting nanoscale systems: {From} the transient to the
  steady-state regime},\ }\href
  {http://link.aps.org/doi/10.1103/PhysRevB.80.115107} {\bibfield  {journal}
  {\bibinfo  {journal} {Phys. Rev. B}\ }\textbf {\bibinfo {volume} {80}},\
  \bibinfo {pages} {115107} (\bibinfo {year} {2009})}\BibitemShut {NoStop}%
\bibitem [{\citenamefont {von Friesen}\ \emph {et~al.}(2009)\citenamefont {von
  Friesen}, \citenamefont {Verdozzi},\ and\ \citenamefont
  {Almbladh}}]{von_friesen_successes_2009}%
  \BibitemOpen
  \bibfield  {author} {\bibinfo {author} {\bibfnamefont {M.~P.}\ \bibnamefont
  {von Friesen}}, \bibinfo {author} {\bibfnamefont {C.}~\bibnamefont
  {Verdozzi}},\ and\ \bibinfo {author} {\bibfnamefont {C.-O.}\ \bibnamefont
  {Almbladh}},\ }\bibfield  {title} {\bibinfo {title} {Successes and {Failures}
  of {Kadanoff}-{Baym} {Dynamics} in {Hubbard} {Nanoclusters}},\ }\href
  {http://link.aps.org/doi/10.1103/PhysRevLett.103.176404} {\bibfield
  {journal} {\bibinfo  {journal} {Phys. Rev. Lett.}\ }\textbf {\bibinfo
  {volume} {103}},\ \bibinfo {pages} {176404} (\bibinfo {year}
  {2009})}\BibitemShut {NoStop}%
\bibitem [{\citenamefont {Sch\"{u}ler}\ \emph {et~al.}(2016)\citenamefont
  {Sch\"{u}ler}, \citenamefont {Berakdar},\ and\ \citenamefont
  {Pavlyukh}}]{schuler_time-dependent_2016}%
  \BibitemOpen
  \bibfield  {author} {\bibinfo {author} {\bibfnamefont {M.}~\bibnamefont
  {Sch\"{u}ler}}, \bibinfo {author} {\bibfnamefont {J.}~\bibnamefont
  {Berakdar}},\ and\ \bibinfo {author} {\bibfnamefont {Y.}~\bibnamefont
  {Pavlyukh}},\ }\bibfield  {title} {\bibinfo {title} {Time-dependent many-body
  treatment of electron-boson dynamics: {Application} to plasmon-accompanied
  photoemission},\ }\href {https://doi.org/10.1103/PhysRevB.93.054303}
  {\bibfield  {journal} {\bibinfo  {journal} {Phys. Rev. B}\ }\textbf {\bibinfo
  {volume} {93}},\ \bibinfo {pages} {054303} (\bibinfo {year}
  {2016})}\BibitemShut {NoStop}%
\bibitem [{\citenamefont {Bittner}\ \emph {et~al.}(2018)\citenamefont
  {Bittner}, \citenamefont {Gole\v{z}}, \citenamefont {Strand}, \citenamefont
  {Eckstein},\ and\ \citenamefont {Werner}}]{bittner_coupled_2018}%
  \BibitemOpen
  \bibfield  {author} {\bibinfo {author} {\bibfnamefont {N.}~\bibnamefont
  {Bittner}}, \bibinfo {author} {\bibfnamefont {D.}~\bibnamefont {Gole\v{z}}},
  \bibinfo {author} {\bibfnamefont {H.~U.~R.}\ \bibnamefont {Strand}}, \bibinfo
  {author} {\bibfnamefont {M.}~\bibnamefont {Eckstein}},\ and\ \bibinfo
  {author} {\bibfnamefont {P.}~\bibnamefont {Werner}},\ }\bibfield  {title}
  {\bibinfo {title} {Coupled charge and spin dynamics in a photoexcited doped
  {Mott} insulator},\ }\href {https://doi.org/10.1103/PhysRevB.97.235125}
  {\bibfield  {journal} {\bibinfo  {journal} {Phys. Rev. B}\ }\textbf {\bibinfo
  {volume} {97}},\ \bibinfo {pages} {235125} (\bibinfo {year}
  {2018})}\BibitemShut {NoStop}%
\bibitem [{\citenamefont {Lipavský}\ \emph {et~al.}(1986)\citenamefont
  {Lipavský}, \citenamefont {\v{S}pi\v{c}ka},\ and\ \citenamefont
  {Velický}}]{lipavsky_generalized_1986}%
  \BibitemOpen
  \bibfield  {author} {\bibinfo {author} {\bibfnamefont {P.}~\bibnamefont
  {Lipavský}}, \bibinfo {author} {\bibfnamefont {V.}~\bibnamefont
  {\v{S}pi\v{c}ka}},\ and\ \bibinfo {author} {\bibfnamefont {B.}~\bibnamefont
  {Velický}},\ }\bibfield  {title} {\bibinfo {title} {Generalized
  {Kadanoff}-{Baym} ansatz for deriving quantum transport equations},\ }\href
  {https://doi.org/10.1103/PhysRevB.34.6933} {\bibfield  {journal} {\bibinfo
  {journal} {Phys. Rev. B}\ }\textbf {\bibinfo {volume} {34}},\ \bibinfo
  {pages} {6933} (\bibinfo {year} {1986})}\BibitemShut {NoStop}%
\bibitem [{\citenamefont {Hermanns}\ \emph {et~al.}(2014)\citenamefont
  {Hermanns}, \citenamefont {Schl\"{u}nzen},\ and\ \citenamefont
  {Bonitz}}]{hermanns_hubbard_2014}%
  \BibitemOpen
  \bibfield  {author} {\bibinfo {author} {\bibfnamefont {S.}~\bibnamefont
  {Hermanns}}, \bibinfo {author} {\bibfnamefont {N.}~\bibnamefont
  {Schl\"{u}nzen}},\ and\ \bibinfo {author} {\bibfnamefont {M.}~\bibnamefont
  {Bonitz}},\ }\bibfield  {title} {\bibinfo {title} {Hubbard nanoclusters far
  from equilibrium},\ }\href {https://doi.org/10.1103/PhysRevB.90.125111}
  {\bibfield  {journal} {\bibinfo  {journal} {Phys. Rev. B}\ }\textbf {\bibinfo
  {volume} {90}},\ \bibinfo {pages} {125111} (\bibinfo {year}
  {2014})}\BibitemShut {NoStop}%
\bibitem [{\citenamefont {Schl\"{u}nzen}\ \emph {et~al.}(2016)\citenamefont
  {Schl\"{u}nzen}, \citenamefont {Hermanns}, \citenamefont {Bonitz},\ and\
  \citenamefont {Verdozzi}}]{schlunzen_dynamics_2016}%
  \BibitemOpen
  \bibfield  {author} {\bibinfo {author} {\bibfnamefont {N.}~\bibnamefont
  {Schl\"{u}nzen}}, \bibinfo {author} {\bibfnamefont {S.}~\bibnamefont
  {Hermanns}}, \bibinfo {author} {\bibfnamefont {M.}~\bibnamefont {Bonitz}},\
  and\ \bibinfo {author} {\bibfnamefont {C.}~\bibnamefont {Verdozzi}},\
  }\bibfield  {title} {\bibinfo {title} {Dynamics of strongly correlated
  fermions: \textit{{Ab} initio} results for two and three dimensions},\ }\href
  {https://doi.org/10.1103/PhysRevB.93.035107} {\bibfield  {journal} {\bibinfo
  {journal} {Phys. Rev. B}\ }\textbf {\bibinfo {volume} {93}},\ \bibinfo
  {pages} {035107} (\bibinfo {year} {2016})}\BibitemShut {NoStop}%
\bibitem [{\citenamefont {Bar~Lev}\ and\ \citenamefont
  {Reichman}(2014)}]{bar_lev_dynamics_2014}%
  \BibitemOpen
  \bibfield  {author} {\bibinfo {author} {\bibfnamefont {Y.}~\bibnamefont
  {Bar~Lev}}\ and\ \bibinfo {author} {\bibfnamefont {D.~R.}\ \bibnamefont
  {Reichman}},\ }\bibfield  {title} {\bibinfo {title} {Dynamics of many-body
  localization},\ }\href {https://doi.org/10.1103/PhysRevB.89.220201}
  {\bibfield  {journal} {\bibinfo  {journal} {Phys. Rev. B}\ }\textbf {\bibinfo
  {volume} {89}},\ \bibinfo {pages} {220201} (\bibinfo {year}
  {2014})}\BibitemShut {NoStop}%
\bibitem [{\citenamefont {Latini}\ \emph {et~al.}(2014)\citenamefont {Latini},
  \citenamefont {Perfetto}, \citenamefont {Uimonen}, \citenamefont {van
  Leeuwen},\ and\ \citenamefont {Stefanucci}}]{latini_charge_2014}%
  \BibitemOpen
  \bibfield  {author} {\bibinfo {author} {\bibfnamefont {S.}~\bibnamefont
  {Latini}}, \bibinfo {author} {\bibfnamefont {E.}~\bibnamefont {Perfetto}},
  \bibinfo {author} {\bibfnamefont {A.-M.}\ \bibnamefont {Uimonen}}, \bibinfo
  {author} {\bibfnamefont {R.}~\bibnamefont {van Leeuwen}},\ and\ \bibinfo
  {author} {\bibfnamefont {G.}~\bibnamefont {Stefanucci}},\ }\bibfield  {title}
  {\bibinfo {title} {Charge dynamics in molecular junctions: {Nonequilibrium}
  {Green}'s function approach made fast},\ }\href
  {https://doi.org/10.1103/PhysRevB.89.075306} {\bibfield  {journal} {\bibinfo
  {journal} {Phys. Rev. B}\ }\textbf {\bibinfo {volume} {89}},\ \bibinfo
  {pages} {075306} (\bibinfo {year} {2014})}\BibitemShut {NoStop}%
\bibitem [{\citenamefont {Perfetto}\ \emph {et~al.}(2015)\citenamefont
  {Perfetto}, \citenamefont {Uimonen}, \citenamefont {van Leeuwen},\ and\
  \citenamefont {Stefanucci}}]{perfetto_first-principles_2015}%
  \BibitemOpen
  \bibfield  {author} {\bibinfo {author} {\bibfnamefont {E.}~\bibnamefont
  {Perfetto}}, \bibinfo {author} {\bibfnamefont {A.-M.}\ \bibnamefont
  {Uimonen}}, \bibinfo {author} {\bibfnamefont {R.}~\bibnamefont {van
  Leeuwen}},\ and\ \bibinfo {author} {\bibfnamefont {G.}~\bibnamefont
  {Stefanucci}},\ }\bibfield  {title} {\bibinfo {title} {First-principles
  nonequilibrium {Green}'s-function approach to transient photoabsorption:
  {Application} to atoms},\ }\href {https://doi.org/10.1103/PhysRevA.92.033419}
  {\bibfield  {journal} {\bibinfo  {journal} {Phys. Rev. A}\ }\textbf {\bibinfo
  {volume} {92}},\ \bibinfo {pages} {033419} (\bibinfo {year}
  {2015})}\BibitemShut {NoStop}%
\bibitem [{\citenamefont {Karlsson}\ \emph {et~al.}(2018)\citenamefont
  {Karlsson}, \citenamefont {van Leeuwen}, \citenamefont {Perfetto},\ and\
  \citenamefont {Stefanucci}}]{karlsson_generalized_2018}%
  \BibitemOpen
  \bibfield  {author} {\bibinfo {author} {\bibfnamefont {D.}~\bibnamefont
  {Karlsson}}, \bibinfo {author} {\bibfnamefont {R.}~\bibnamefont {van
  Leeuwen}}, \bibinfo {author} {\bibfnamefont {E.}~\bibnamefont {Perfetto}},\
  and\ \bibinfo {author} {\bibfnamefont {G.}~\bibnamefont {Stefanucci}},\
  }\bibfield  {title} {\bibinfo {title} {The generalized {Kadanoff}-{Baym}
  ansatz with initial correlations},\ }\href
  {https://doi.org/10.1103/PhysRevB.98.115148} {\bibfield  {journal} {\bibinfo
  {journal} {Phys. Rev. B}\ }\textbf {\bibinfo {volume} {98}},\ \bibinfo
  {pages} {115148} (\bibinfo {year} {2018})}\BibitemShut {NoStop}%
\bibitem [{\citenamefont {Perfetto}\ \emph {et~al.}(2018)\citenamefont
  {Perfetto}, \citenamefont {Sangalli}, \citenamefont {Marini},\ and\
  \citenamefont {Stefanucci}}]{perfetto_ultrafast_2018}%
  \BibitemOpen
  \bibfield  {author} {\bibinfo {author} {\bibfnamefont {E.}~\bibnamefont
  {Perfetto}}, \bibinfo {author} {\bibfnamefont {D.}~\bibnamefont {Sangalli}},
  \bibinfo {author} {\bibfnamefont {A.}~\bibnamefont {Marini}},\ and\ \bibinfo
  {author} {\bibfnamefont {G.}~\bibnamefont {Stefanucci}},\ }\bibfield  {title}
  {\bibinfo {title} {Ultrafast {Charge} {Migration} in {XUV} {Photoexcited}
  {Phenylalanine}: {A} {First}-{Principles} {Study} {Based} on {Real}-{Time}
  {Nonequilibrium} {Green}'s {Functions}},\ }\href
  {https://doi.org/10.1021/acs.jpclett.8b00025} {\bibfield  {journal} {\bibinfo
   {journal} {J. Phys. Chem. Lett.}\ }\textbf {\bibinfo {volume} {9}},\
  \bibinfo {pages} {1353} (\bibinfo {year} {2018})}\BibitemShut {NoStop}%
\bibitem [{\citenamefont {Schl\"{u}nzen}\ \emph {et~al.}(2020)\citenamefont
  {Schl\"{u}nzen}, \citenamefont {Joost},\ and\ \citenamefont
  {Bonitz}}]{schlunzen_achieving_2020}%
  \BibitemOpen
  \bibfield  {author} {\bibinfo {author} {\bibfnamefont {N.}~\bibnamefont
  {Schl\"{u}nzen}}, \bibinfo {author} {\bibfnamefont {J.-P.}\ \bibnamefont
  {Joost}},\ and\ \bibinfo {author} {\bibfnamefont {M.}~\bibnamefont
  {Bonitz}},\ }\bibfield  {title} {\bibinfo {title} {Achieving the {Scaling}
  {Limit} for {Nonequilibrium} {Green} {Functions} {Simulations}},\ }\href
  {https://doi.org/10.1103/PhysRevLett.124.076601} {\bibfield  {journal}
  {\bibinfo  {journal} {Phys. Rev. Lett.}\ }\textbf {\bibinfo {volume} {124}},\
  \bibinfo {pages} {076601} (\bibinfo {year} {2020})}\BibitemShut {NoStop}%
\bibitem [{\citenamefont {Joost}\ \emph {et~al.}(2020)\citenamefont {Joost},
  \citenamefont {Schl\"{u}nzen},\ and\ \citenamefont
  {Bonitz}}]{joost_g1-g2_2020}%
  \BibitemOpen
  \bibfield  {author} {\bibinfo {author} {\bibfnamefont {J.-P.}\ \bibnamefont
  {Joost}}, \bibinfo {author} {\bibfnamefont {N.}~\bibnamefont
  {Schl\"{u}nzen}},\ and\ \bibinfo {author} {\bibfnamefont {M.}~\bibnamefont
  {Bonitz}},\ }\bibfield  {title} {\bibinfo {title} {G1-{G2} scheme: {Dramatic}
  acceleration of nonequilibrium {Green} functions simulations within the
  {Hartree}-{Fock} generalized {Kadanoff}-{Baym} ansatz},\ }\href
  {https://doi.org/10.1103/PhysRevB.101.245101} {\bibfield  {journal} {\bibinfo
   {journal} {Phys. Rev. B}\ }\textbf {\bibinfo {volume} {101}},\ \bibinfo
  {pages} {245101} (\bibinfo {year} {2020})}\BibitemShut {NoStop}%
\bibitem [{\citenamefont {Pavlyukh}\ \emph {et~al.}(2021)\citenamefont
  {Pavlyukh}, \citenamefont {Perfetto},\ and\ \citenamefont
  {Stefanucci}}]{pavlyukh_photoinduced_2021}%
  \BibitemOpen
  \bibfield  {author} {\bibinfo {author} {\bibfnamefont {Y.}~\bibnamefont
  {Pavlyukh}}, \bibinfo {author} {\bibfnamefont {E.}~\bibnamefont {Perfetto}},\
  and\ \bibinfo {author} {\bibfnamefont {G.}~\bibnamefont {Stefanucci}},\
  }\bibfield  {title} {\bibinfo {title} {Photoinduced dynamics of organic
  molecules using nonequilibrium {Green}'s functions with second-{Born},
  \textit{{GW}}, \textit{{T}}-matrix, and three-particle correlations},\ }\href
  {https://doi.org/10.1103/PhysRevB.104.035124} {\bibfield  {journal} {\bibinfo
   {journal} {Phys. Rev. B}\ }\textbf {\bibinfo {volume} {104}},\ \bibinfo
  {pages} {035124} (\bibinfo {year} {2021})}\BibitemShut {NoStop}%
\bibitem [{\citenamefont {Perfetto}\ \emph {et~al.}(2021)\citenamefont
  {Perfetto}, \citenamefont {Pavlyukh},\ and\ \citenamefont
  {Stefanucci}}]{perfetto_tdgw_2021}%
  \BibitemOpen
  \bibfield  {author} {\bibinfo {author} {\bibfnamefont {E.}~\bibnamefont
  {Perfetto}}, \bibinfo {author} {\bibfnamefont {Y.}~\bibnamefont {Pavlyukh}},\
  and\ \bibinfo {author} {\bibfnamefont {G.}~\bibnamefont {Stefanucci}},\
  }\bibfield  {title} {\bibinfo {title} {Real-time {$GW$}: \emph{ab initio}
  description of the ultrafast carrier and exciton dynamics in two-dimensional
  systems},\ }\href {http://arxiv.org/abs/2109.15209} {\bibfield  {journal}
  {\bibinfo  {journal} {arXiv:2109.15209}\ } (\bibinfo {year}
  {2021})}\BibitemShut {NoStop}%
\bibitem [{\citenamefont {Borkowski}\ \emph {et~al.}(2021)\citenamefont
  {Borkowski}, \citenamefont {Schl\"unzen}, \citenamefont {Joost},
  \citenamefont {Reiser},\ and\ \citenamefont
  {Bonitz}}]{Borkowski_doublon_2021}%
  \BibitemOpen
  \bibfield  {author} {\bibinfo {author} {\bibfnamefont {L.}~\bibnamefont
  {Borkowski}}, \bibinfo {author} {\bibfnamefont {N.}~\bibnamefont
  {Schl\"unzen}}, \bibinfo {author} {\bibfnamefont {J.~P.}\ \bibnamefont
  {Joost}}, \bibinfo {author} {\bibfnamefont {F.}~\bibnamefont {Reiser}},\ and\
  \bibinfo {author} {\bibfnamefont {M.}~\bibnamefont {Bonitz}},\ }\bibfield
  {title} {\bibinfo {title} {Doublon production in correlated materials by
  multiple ion impacts},\ }\href {https://arxiv.org/abs/2110.06644} {\bibfield
  {journal} {\bibinfo  {journal} {arXiv:2110.06644}\ } (\bibinfo {year}
  {2021})}\BibitemShut {NoStop}%
\bibitem [{\citenamefont {Baym}\ and\ \citenamefont
  {Kadanoff}(1961)}]{baym_conservation_1961}%
  \BibitemOpen
  \bibfield  {author} {\bibinfo {author} {\bibfnamefont {G.}~\bibnamefont
  {Baym}}\ and\ \bibinfo {author} {\bibfnamefont {L.~P.}\ \bibnamefont
  {Kadanoff}},\ }\bibfield  {title} {\bibinfo {title} {Conservation {Laws} and
  {Correlation} {Functions}},\ }\href {https://doi.org/10.1103/PhysRev.124.287}
  {\bibfield  {journal} {\bibinfo  {journal} {Phys. Rev.}\ }\textbf {\bibinfo
  {volume} {124}},\ \bibinfo {pages} {287} (\bibinfo {year}
  {1961})}\BibitemShut {NoStop}%
\bibitem [{\citenamefont {Baym}(1962)}]{baym_self-consistent_1962}%
  \BibitemOpen
  \bibfield  {author} {\bibinfo {author} {\bibfnamefont {G.}~\bibnamefont
  {Baym}},\ }\bibfield  {title} {\bibinfo {title} {Self-{Consistent}
  {Approximations} in {Many}-{Body} {Systems}},\ }\href
  {https://doi.org/10.1103/PhysRev.127.1391} {\bibfield  {journal} {\bibinfo
  {journal} {Phys. Rev.}\ }\textbf {\bibinfo {volume} {127}},\ \bibinfo {pages}
  {1391} (\bibinfo {year} {1962})}\BibitemShut {NoStop}%
\bibitem [{\citenamefont {S\"akkinen}(2016)}]{nikothesis}%
  \BibitemOpen
  \bibfield  {author} {\bibinfo {author} {\bibfnamefont {N.}~\bibnamefont
  {S\"akkinen}},\ }\emph {\bibinfo {title} {Application of time-dependent
  many-body perturbation theory to excitation spectra of selected finite model
  systems}},\ \href {https://jyx.jyu.fi/handle/123456789/53076} {Ph.D.
  thesis},\ \bibinfo  {school} {University of Jyv\"askyl\"a} (\bibinfo {year}
  {2016})\BibitemShut {NoStop}%
\bibitem [{\citenamefont {Pellegrini}\ \emph {et~al.}(2015)\citenamefont
  {Pellegrini}, \citenamefont {Flick}, \citenamefont {Tokatly}, \citenamefont
  {Appel},\ and\ \citenamefont {Rubio}}]{pellegrini_optimized_2015}%
  \BibitemOpen
  \bibfield  {author} {\bibinfo {author} {\bibfnamefont {C.}~\bibnamefont
  {Pellegrini}}, \bibinfo {author} {\bibfnamefont {J.}~\bibnamefont {Flick}},
  \bibinfo {author} {\bibfnamefont {I.~V.}\ \bibnamefont {Tokatly}}, \bibinfo
  {author} {\bibfnamefont {H.}~\bibnamefont {Appel}},\ and\ \bibinfo {author}
  {\bibfnamefont {A.}~\bibnamefont {Rubio}},\ }\bibfield  {title} {\bibinfo
  {title} {Optimized {Effective} {Potential} for {Quantum} {Electrodynamical}
  {Time}-{Dependent} {Density} {Functional} {Theory}},\ }\href
  {https://doi.org/10.1103/PhysRevLett.115.093001} {\bibfield  {journal}
  {\bibinfo  {journal} {Phys. Rev. Lett.}\ }\textbf {\bibinfo {volume} {115}},\
  \bibinfo {pages} {093001} (\bibinfo {year} {2015})}\BibitemShut {NoStop}%
\bibitem [{\citenamefont {Fan}(1951)}]{fan_temperature_1951}%
  \BibitemOpen
  \bibfield  {author} {\bibinfo {author} {\bibfnamefont {H.~Y.}\ \bibnamefont
  {Fan}},\ }\bibfield  {title} {\bibinfo {title} {Temperature {Dependence} of
  the {Energy} {Gap} in {Semiconductors}},\ }\href
  {https://doi.org/10.1103/PhysRev.82.900} {\bibfield  {journal} {\bibinfo
  {journal} {Phys. Rev.}\ }\textbf {\bibinfo {volume} {82}},\ \bibinfo {pages}
  {900} (\bibinfo {year} {1951})}\BibitemShut {NoStop}%
\bibitem [{\citenamefont {Marini}(2008)}]{marini_ab_2008}%
  \BibitemOpen
  \bibfield  {author} {\bibinfo {author} {\bibfnamefont {A.}~\bibnamefont
  {Marini}},\ }\bibfield  {title} {\bibinfo {title} {\textit{{Ab} {Initio}}
  {Finite}-{Temperature} {Excitons}},\ }\href
  {https://doi.org/10.1103/PhysRevLett.101.106405} {\bibfield  {journal}
  {\bibinfo  {journal} {Phys. Rev. Lett.}\ }\textbf {\bibinfo {volume} {101}},\
  \bibinfo {pages} {106405} (\bibinfo {year} {2008})}\BibitemShut {NoStop}%
\bibitem [{\citenamefont {Marini}\ and\ \citenamefont
  {Del~Sole}(2003)}]{marini_dynamical_2003}%
  \BibitemOpen
  \bibfield  {author} {\bibinfo {author} {\bibfnamefont {A.}~\bibnamefont
  {Marini}}\ and\ \bibinfo {author} {\bibfnamefont {R.}~\bibnamefont
  {Del~Sole}},\ }\bibfield  {title} {\bibinfo {title} {Dynamical {Excitonic}
  {Effects} in {Metals} and {Semiconductors}},\ }\href
  {https://doi.org/10.1103/PhysRevLett.91.176402} {\bibfield  {journal}
  {\bibinfo  {journal} {Phys. Rev. Lett.}\ }\textbf {\bibinfo {volume} {91}},\
  \bibinfo {pages} {176402} (\bibinfo {year} {2003})}\BibitemShut {NoStop}%
\bibitem [{\citenamefont {Haug}(1992)}]{haug_interband_1992}%
  \BibitemOpen
  \bibfield  {author} {\bibinfo {author} {\bibfnamefont {H.}~\bibnamefont
  {Haug}},\ }\bibfield  {title} {\bibinfo {title} {Interband {Quantum}
  {Kinetics} with {LO}-{Phonon} {Scattering} in a {Laser}-{Pulse}-{Excited}
  {Semiconductor} {I}. {Theory}},\ }\href
  {https://doi.org/10.1002/pssb.2221730114} {\bibfield  {journal} {\bibinfo
  {journal} {Phys. Status Solidi B}\ }\textbf {\bibinfo {volume} {173}},\
  \bibinfo {pages} {139} (\bibinfo {year} {1992})}\BibitemShut {NoStop}%
\bibitem [{\citenamefont {Bonitz}\ \emph {et~al.}(1999)\citenamefont {Bonitz},
  \citenamefont {Semkat},\ and\ \citenamefont
  {Haug}}]{bonitz_non-lorentzian_1999}%
  \BibitemOpen
  \bibfield  {author} {\bibinfo {author} {\bibfnamefont {M.}~\bibnamefont
  {Bonitz}}, \bibinfo {author} {\bibfnamefont {D.}~\bibnamefont {Semkat}},\
  and\ \bibinfo {author} {\bibfnamefont {H.}~\bibnamefont {Haug}},\ }\bibfield
  {title} {\bibinfo {title} {Non-{Lorentzian} spectral functions for {Coulomb}
  quantum kinetics},\ }\href {https://doi.org/10.1007/s100510050770} {\bibfield
   {journal} {\bibinfo  {journal} {Eur. Phys. J. B}\ }\textbf {\bibinfo
  {volume} {9}},\ \bibinfo {pages} {309} (\bibinfo {year} {1999})}\BibitemShut
  {NoStop}%
\bibitem [{\citenamefont {Ren}\ \emph {et~al.}(2015)\citenamefont {Ren},
  \citenamefont {Marom}, \citenamefont {Caruso}, \citenamefont {Scheffler},\
  and\ \citenamefont {Rinke}}]{ren_beyond_2015}%
  \BibitemOpen
  \bibfield  {author} {\bibinfo {author} {\bibfnamefont {X.}~\bibnamefont
  {Ren}}, \bibinfo {author} {\bibfnamefont {N.}~\bibnamefont {Marom}}, \bibinfo
  {author} {\bibfnamefont {F.}~\bibnamefont {Caruso}}, \bibinfo {author}
  {\bibfnamefont {M.}~\bibnamefont {Scheffler}},\ and\ \bibinfo {author}
  {\bibfnamefont {P.}~\bibnamefont {Rinke}},\ }\bibfield  {title} {\bibinfo
  {title} {Beyond the \textit{{GW}} approximation: {A} second-order screened
  exchange correction},\ }\href {https://doi.org/10.1103/PhysRevB.92.081104}
  {\bibfield  {journal} {\bibinfo  {journal} {Phys. Rev. B}\ }\textbf {\bibinfo
  {volume} {92}},\ \bibinfo {pages} {081104(R)} (\bibinfo {year}
  {2015})}\BibitemShut {NoStop}%
\bibitem [{\citenamefont {Maggio}\ and\ \citenamefont
  {Kresse}(2017)}]{maggio_gw_2017}%
  \BibitemOpen
  \bibfield  {author} {\bibinfo {author} {\bibfnamefont {E.}~\bibnamefont
  {Maggio}}\ and\ \bibinfo {author} {\bibfnamefont {G.}~\bibnamefont
  {Kresse}},\ }\bibfield  {title} {\bibinfo {title} {\textit{{GW}} {Vertex}
  {Corrected} {Calculations} for {Molecular} {Systems}},\ }\href
  {https://doi.org/10.1021/acs.jctc.7b00586} {\bibfield  {journal} {\bibinfo
  {journal} {J. Chem. Theory Comput.}\ }\textbf {\bibinfo {volume} {13}},\
  \bibinfo {pages} {4765} (\bibinfo {year} {2017})}\BibitemShut {NoStop}%
\bibitem [{\citenamefont {Pavlyukh}\ \emph {et~al.}(2020)\citenamefont
  {Pavlyukh}, \citenamefont {Stefanucci},\ and\ \citenamefont {van
  Leeuwen}}]{pavlyukh_dynamically_2020}%
  \BibitemOpen
  \bibfield  {author} {\bibinfo {author} {\bibfnamefont {Y.}~\bibnamefont
  {Pavlyukh}}, \bibinfo {author} {\bibfnamefont {G.}~\bibnamefont
  {Stefanucci}},\ and\ \bibinfo {author} {\bibfnamefont {R.}~\bibnamefont {van
  Leeuwen}},\ }\bibfield  {title} {\bibinfo {title} {Dynamically screened
  vertex correction to \textit{{GW}}},\ }\href
  {https://doi.org/10.1103/PhysRevB.102.045121} {\bibfield  {journal} {\bibinfo
   {journal} {Phys. Rev. B}\ }\textbf {\bibinfo {volume} {102}},\ \bibinfo
  {pages} {045121} (\bibinfo {year} {2020})}\BibitemShut {NoStop}%
\bibitem [{\citenamefont {Wang}\ \emph {et~al.}(2021)\citenamefont {Wang},
  \citenamefont {Rinke},\ and\ \citenamefont {Ren}}]{wang_assessing_2021}%
  \BibitemOpen
  \bibfield  {author} {\bibinfo {author} {\bibfnamefont {Y.}~\bibnamefont
  {Wang}}, \bibinfo {author} {\bibfnamefont {P.}~\bibnamefont {Rinke}},\ and\
  \bibinfo {author} {\bibfnamefont {X.}~\bibnamefont {Ren}},\ }\bibfield
  {title} {\bibinfo {title} {Assessing the
  \textit{{G}}$_{\textrm{0}}$\textit{{W}}$_{\textrm{0}}$$\gamma$$_{\textrm{0}}$
  $^{\textrm{(1)}}$ {Approach}: {Beyond}
  \textit{{G}}$_{\textrm{0}}$\textit{{W}}$_{\textrm{0}}$ with {Hedin}'s {Full}
  {Second}-{Order} {Self}-{Energy} {Contribution}},\ }\href
  {https://doi.org/10.1021/acs.jctc.1c00488} {\bibfield  {journal} {\bibinfo
  {journal} {J. Chem. Theory Comput.}\ }\textbf {\bibinfo {volume} {17}},\
  \bibinfo {pages} {5140} (\bibinfo {year} {2021})}\BibitemShut {NoStop}%
\bibitem [{\citenamefont {Tuovinen}\ \emph {et~al.}(2019)\citenamefont
  {Tuovinen}, \citenamefont {Gole\v{z}}, \citenamefont {Sch\"{u}ler},
  \citenamefont {Werner}, \citenamefont {Eckstein},\ and\ \citenamefont
  {Sentef}}]{tuovinen_adiabatic_2019}%
  \BibitemOpen
  \bibfield  {author} {\bibinfo {author} {\bibfnamefont {R.}~\bibnamefont
  {Tuovinen}}, \bibinfo {author} {\bibfnamefont {D.}~\bibnamefont {Gole\v{z}}},
  \bibinfo {author} {\bibfnamefont {M.}~\bibnamefont {Sch\"{u}ler}}, \bibinfo
  {author} {\bibfnamefont {P.}~\bibnamefont {Werner}}, \bibinfo {author}
  {\bibfnamefont {M.}~\bibnamefont {Eckstein}},\ and\ \bibinfo {author}
  {\bibfnamefont {M.~A.}\ \bibnamefont {Sentef}},\ }\bibfield  {title}
  {\bibinfo {title} {Adiabatic {Preparation} of a {Correlated}
  {Symmetry}‐{Broken} {Initial} {State} with the {Generalized}
  {Kadanoff}-{Baym} {Ansatz}},\ }\href {https://doi.org/10.1002/pssb.201800469}
  {\bibfield  {journal} {\bibinfo  {journal} {Phys. Status Solidi B}\ }\textbf
  {\bibinfo {volume} {256}},\ \bibinfo {pages} {1800469} (\bibinfo {year}
  {2019})}\BibitemShut {NoStop}%
\bibitem [{\citenamefont {Tuovinen}\ \emph {et~al.}(2020)\citenamefont
  {Tuovinen}, \citenamefont {Gole\v{z}}, \citenamefont {Eckstein},\ and\
  \citenamefont {Sentef}}]{tuovinen_comparing_2020}%
  \BibitemOpen
  \bibfield  {author} {\bibinfo {author} {\bibfnamefont {R.}~\bibnamefont
  {Tuovinen}}, \bibinfo {author} {\bibfnamefont {D.}~\bibnamefont {Gole\v{z}}},
  \bibinfo {author} {\bibfnamefont {M.}~\bibnamefont {Eckstein}},\ and\
  \bibinfo {author} {\bibfnamefont {M.~A.}\ \bibnamefont {Sentef}},\ }\bibfield
   {title} {\bibinfo {title} {Comparing the generalized {Kadanoff}-{Baym}
  ansatz with the full {Kadanoff}-{Baym} equations for an excitonic insulator
  out of equilibrium},\ }\href {https://doi.org/10.1103/PhysRevB.102.115157}
  {\bibfield  {journal} {\bibinfo  {journal} {Phys. Rev. B}\ }\textbf {\bibinfo
  {volume} {102}},\ \bibinfo {pages} {115157} (\bibinfo {year}
  {2020})}\BibitemShut {NoStop}%
\bibitem [{\citenamefont {Fehske}\ \emph {et~al.}(2011)\citenamefont {Fehske},
  \citenamefont {Wellein},\ and\ \citenamefont
  {Bishop}}]{fehske_spatiotemporal_2011}%
  \BibitemOpen
  \bibfield  {author} {\bibinfo {author} {\bibfnamefont {H.}~\bibnamefont
  {Fehske}}, \bibinfo {author} {\bibfnamefont {G.}~\bibnamefont {Wellein}},\
  and\ \bibinfo {author} {\bibfnamefont {A.~R.}\ \bibnamefont {Bishop}},\
  }\bibfield  {title} {\bibinfo {title} {Spatiotemporal evolution of polaronic
  states in finite quantum systems},\ }\href
  {https://doi.org/10.1103/PhysRevB.83.075104} {\bibfield  {journal} {\bibinfo
  {journal} {Phys. Rev. B}\ }\textbf {\bibinfo {volume} {83}},\ \bibinfo
  {pages} {075104} (\bibinfo {year} {2011})}\BibitemShut {NoStop}%
\bibitem [{\citenamefont {Perfetto}\ \emph {et~al.}(2020)\citenamefont
  {Perfetto}, \citenamefont {Trabattoni}, \citenamefont {Calegari},
  \citenamefont {Nisoli}, \citenamefont {Marini},\ and\ \citenamefont
  {Stefanucci}}]{perfetto_ultrafast_2020}%
  \BibitemOpen
  \bibfield  {author} {\bibinfo {author} {\bibfnamefont {E.}~\bibnamefont
  {Perfetto}}, \bibinfo {author} {\bibfnamefont {A.}~\bibnamefont
  {Trabattoni}}, \bibinfo {author} {\bibfnamefont {F.}~\bibnamefont
  {Calegari}}, \bibinfo {author} {\bibfnamefont {M.}~\bibnamefont {Nisoli}},
  \bibinfo {author} {\bibfnamefont {A.}~\bibnamefont {Marini}},\ and\ \bibinfo
  {author} {\bibfnamefont {G.}~\bibnamefont {Stefanucci}},\ }\bibfield  {title}
  {\bibinfo {title} {Ultrafast {Quantum} {Interference} in the {Charge}
  {Migration} of {Tryptophan}},\ }\href@noop {} {\bibfield  {journal} {\bibinfo
   {journal} {J. Phys. Chem. Lett.}\ ,\ \bibinfo {pages} {9}} (\bibinfo {year}
  {2020})}\BibitemShut {NoStop}%
\bibitem [{\citenamefont {M{\aa}nsson}\ \emph {et~al.}(2021)\citenamefont
  {M{\aa}nsson}, \citenamefont {Latini}, \citenamefont {Covito}, \citenamefont
  {Wanie}, \citenamefont {Galli}, \citenamefont {Perfetto}, \citenamefont
  {Stefanucci}, \citenamefont {H{\"u}bener}, \citenamefont {De~Giovannini},
  \citenamefont {Castrovilli}, \citenamefont {Trabattoni}, \citenamefont
  {Frassetto}, \citenamefont {Poletto}, \citenamefont {Greenwood},
  \citenamefont {L{\'e}gar{\'e}}, \citenamefont {Nisoli}, \citenamefont
  {Rubio},\ and\ \citenamefont {Calegari}}]{Adenine_2021}%
  \BibitemOpen
  \bibfield  {author} {\bibinfo {author} {\bibfnamefont {E.~P.}\ \bibnamefont
  {M{\aa}nsson}}, \bibinfo {author} {\bibfnamefont {S.}~\bibnamefont {Latini}},
  \bibinfo {author} {\bibfnamefont {F.}~\bibnamefont {Covito}}, \bibinfo
  {author} {\bibfnamefont {V.}~\bibnamefont {Wanie}}, \bibinfo {author}
  {\bibfnamefont {M.}~\bibnamefont {Galli}}, \bibinfo {author} {\bibfnamefont
  {E.}~\bibnamefont {Perfetto}}, \bibinfo {author} {\bibfnamefont
  {G.}~\bibnamefont {Stefanucci}}, \bibinfo {author} {\bibfnamefont
  {H.}~\bibnamefont {H{\"u}bener}}, \bibinfo {author} {\bibfnamefont
  {U.}~\bibnamefont {De~Giovannini}}, \bibinfo {author} {\bibfnamefont {M.~C.}\
  \bibnamefont {Castrovilli}}, \bibinfo {author} {\bibfnamefont
  {A.}~\bibnamefont {Trabattoni}}, \bibinfo {author} {\bibfnamefont
  {F.}~\bibnamefont {Frassetto}}, \bibinfo {author} {\bibfnamefont
  {L.}~\bibnamefont {Poletto}}, \bibinfo {author} {\bibfnamefont {J.~B.}\
  \bibnamefont {Greenwood}}, \bibinfo {author} {\bibfnamefont {F.}~\bibnamefont
  {L{\'e}gar{\'e}}}, \bibinfo {author} {\bibfnamefont {M.}~\bibnamefont
  {Nisoli}}, \bibinfo {author} {\bibfnamefont {A.}~\bibnamefont {Rubio}},\ and\
  \bibinfo {author} {\bibfnamefont {F.}~\bibnamefont {Calegari}},\ }\bibfield
  {title} {\bibinfo {title} {Real-time observation of a correlation-driven sub
  3{\thinspace}fs charge migration in ionised adenine},\ }\href
  {https://doi.org/10.1038/s42004-021-00510-5} {\bibfield  {journal} {\bibinfo
  {journal} {Communications Chemistry}\ }\textbf {\bibinfo {volume} {4}},\
  \bibinfo {pages} {73} (\bibinfo {year} {2021})}\BibitemShut {NoStop}%
\bibitem [{\citenamefont {Tuovinen}\ \emph {et~al.}(2021)\citenamefont
  {Tuovinen}, \citenamefont {van Leeuwen}, \citenamefont {Perfetto},\ and\
  \citenamefont {Stefanucci}}]{tuovinen_electronic_2021}%
  \BibitemOpen
  \bibfield  {author} {\bibinfo {author} {\bibfnamefont {R.}~\bibnamefont
  {Tuovinen}}, \bibinfo {author} {\bibfnamefont {R.}~\bibnamefont {van
  Leeuwen}}, \bibinfo {author} {\bibfnamefont {E.}~\bibnamefont {Perfetto}},\
  and\ \bibinfo {author} {\bibfnamefont {G.}~\bibnamefont {Stefanucci}},\
  }\bibfield  {title} {\bibinfo {title} {Electronic transport in molecular
  junctions: {The} generalized {Kadanoff}-{Baym} ansatz with initial contact
  and correlations},\ }\href {https://doi.org/10.1063/5.0040685} {\bibfield
  {journal} {\bibinfo  {journal} {J. Chem. Phys.}\ }\textbf {\bibinfo {volume}
  {154}},\ \bibinfo {pages} {094104} (\bibinfo {year} {2021})}\BibitemShut
  {NoStop}%
\end{thebibliography}
%

\end{document}